\documentclass{emulateapj}

\usepackage{amssymb}
\usepackage{amsmath}
\usepackage{graphicx}
\usepackage{natbib}
\usepackage{txfonts}
\usepackage{bbold}
\usepackage{array}
\usepackage[colorlinks=true,citecolor=blue,linkcolor=blue]{hyperref}

\newcommand\T{\rule{0pt}{2.9ex}}       
\newcommand\B{\rule[-1.4ex]{0pt}{0pt}} 

\def\roma{1}
\def\pescara{2}
\def\unicamp{3}
\def\rio{4}

\begin{document}

\title{Neutrino oscillations within the induced gravitational collapse paradigm of long gamma-ray bursts}


\author{L.~Becerra\altaffilmark{\roma,\pescara},
		M.~M.~Guzzo\altaffilmark{\unicamp},		
        F. Rossi-Torres\altaffilmark{\unicamp},
        J.~A.~Rueda\altaffilmark{\roma,\pescara,\rio},
        R.~Ruffini\altaffilmark{\roma,\pescara,\rio}
        and
		J.~D.~Uribe\altaffilmark{\roma,\pescara}}
	
\altaffiltext{\roma}{Dipartimento di Fisica and ICRA, 
                     Sapienza Universit\`a di Roma, 
                     P.le Aldo Moro 5, 
                     I--00185 Rome, 
                     Italy}

\altaffiltext{\pescara}{ICRANet, 
                     P.zza della Repubblica 10, 
                     I--65122 Pescara, 
                     Italy}

\altaffiltext{\rio}{ICRANet-Rio, i
                    Centro Brasileiro de Pesquisas F\'isicas, 
                    Rua Dr. Xavier Sigaud 150, 
                    22290--180 Rio de Janeiro, 
                    Brazil}

\altaffiltext{\unicamp}{Instituto de F\'isica Gleb Wataghin,							Universidade Estadual de Campinas, Rua 										S\'ergio Buarque de Holanda 777, 13083-859 									Campinas SP Brazil}

\date{\today}

\begin{abstract}
The induced gravitational collapse (IGC) paradigm of long gamma-ray bursts (GRBs) associated with supernovae (SNe) predicts a copious neutrino-antineutrino ($\nu\bar{\nu}$) emission owing to the hypercritical accretion process of SN ejecta onto a neutron star (NS) binary companion. The neutrino emission can reach luminosities of up to $10^{57}$~MeV~s$^{-1}$, mean neutrino energies 20~MeV, and neutrino densities $10^{31}$~cm$^{-3}$. Along their path from the vicinity of the NS surface outward, such neutrinos experience flavor transformations dictated by the neutrino to electron density ratio. We determine the neutrino and electron on the accretion zone and use them to compute the neutrino flavor evolution. For normal and inverted neutrino-mass hierarchies and within the two-flavor formalism ($\nu_{e}\nu_{x}$), we estimate the final electronic and non-electronic neutrino content after two oscillation processes: (1) neutrino collective effects due to neutrino self-interactions where the neutrino density dominates and, (2) the Mikheyev-Smirnov-Wolfenstein (MSW) effect, where the electron density dominates. We find that the final neutrino content is composed by $\sim$55\% ($\sim$62\%) of electronic neutrinos, i.e. $\nu_{e}+\bar{\nu}_{e}$, for the normal (inverted) neutrino-mass hierarchy. The results of this work are the first step toward the characterization of a novel source of astrophysical MeV-neutrinos in addition to core-collapse SNe and, as such, deserve further attention.
\end{abstract}

\maketitle

\section{Introduction}

The emergent picture of gamma-ray burst (GRB) is that both, short-duration and long-duration GRBs, originate from binary systems \citep{2016ApJ...832..136R}.

Short bursts originate from neutron star-neutron star (NS-NS) or neutron star-black hole (NS-BH) mergers \citep[see, e.g.,][]{1986ApJ...308L..47G,1986ApJ...308L..43P,1989Natur.340..126E,1991ApJ...379L..17N}. For this case \citet{1992ApJ...395L..83N} introduced the role of neutrino-antineutrino ($\nu\bar\nu$) annihilation leading to the formation of an electron-positron plasma ($e^{-}e^{+}$) in NS-NS and NS-BH mergers. Such a result triggered many theoretical works, including the general relativistic treatment by \citet{2002ApJ...578..310S} of the $\nu\bar\nu$ annihilation process giving rise to the $e^{-}e^{+}$ plasma in a NS-NS system.  

For long bursts we stand on the induced gravitational collapse (IGC) paradigm \citep{2006tmgm.meet..369R,2008mgm..conf..368R,2012A&A...548L...5I,2012ApJ...758L...7R,2014ApJ...793L..36F,2015ApJ...798...10R}, based on the hypercritical accretion process of the supernova (SN) ejecta of the explosion of a carbon-oxygen core (CO$_{\rm core}$) onto a NS binary companion. In the above processes, the emission of neutrinos is a key ingredient. 

We focus hereafter on the neutrino emission of long bursts within the IGC scenario. The role of neutrinos in this paradigm has been recently addressed \citet{2014ApJ...793L..36F,2015PhRvL.115w1102F,2015ApJ...812..100B,2016ApJ...833..107B}. The hypercritical accretion of the SN ejecta onto the NS companion can reach very high rates of up to $10^{-2}~M_\odot$~s$^{-1}$ and its duration can be of the order of $10$--$10^4$~s depending on the binary parameters. The photons become trapped within the accretion flow and thus do not serve as an energy sink. The high temperature developed on the NS surface leads to $e^{-}e^{+}$ pairs that, via weak interactions, annihilate into $\nu\bar\nu$ pairs with neutrino luminosities of up to $10^{52}$~erg~s$^{-1}$ for the highest accretion rates. Thus, this process dominates the cooling and give rise to a very efficient conversion of the gravitational energy gained by accretion into radiation. We refer to \citet{2016ApJ...833..107B} for further details on this process.

The above hypercritical accretion process can lead the NS to two alternative fates, leading to the existence of two long GRB sub-classes \citep{2014ApJ...793L..36F,2015PhRvL.115w1102F,2015ApJ...812..100B,2016ApJ...833..107B,2016ApJ...832..136R}:
\begin{itemize}
\item[I.]
The hypercritical accretion leads to a more massive NS companion but not to a black hole (BH). These binaries explain the X-ray flashes (XRFs); long bursts with isotropic energy $E_{\rm iso}\lesssim 10^{52}$~erg and rest-frame spectral peak energy $E_{p,i}\lesssim 200$~keV \citep[see][for further details]{2016ApJ...832..136R}. The local observed number density rate of this GRB sub-class is \citep{2016ApJ...832..136R}: $\rho_{\rm GRB} = 100^{+45}_{-34}$~Gpc$^{-3}$yr$^{-1}$.
\item[II.]
The hypercritical accretion is high enough to make the NS reach its critical mass triggering its gravitational collapse with consequent BH formation. These binaries explain the binary-driven hypernovae (BdHNe); long bursts with $E_{\rm iso}\gtrsim10^{52}$~erg and $E_{p,i}\gtrsim200$~keV \citep[see][for further details]{2016ApJ...832..136R}. The local observed number density rate of this GRB sub-class is \citep{2016ApJ...832..136R}: $\rho_{\rm GRB} = 0.77^{+0.09}_{-0.08}$~Gpc$^{-3}$yr$^{-1}$.
\end{itemize}

Simulations of the hypercritical accretion process in the above binaries have been presented in \citet{2014ApJ...793L..36F,2015PhRvL.115w1102F,2015ApJ...812..100B,2016ApJ...833..107B}. It has been shown how, thanks to the development of a copious neutrino emission near the NS surface, the NS is allowed to accrete matter from the SN at very high rates. The specific conditions leading to XRFs and BdHNe as well as a detailed analysis of the neutrino production in these systems have been presented in \citet{2016ApJ...833..107B}. Neutrino emission can reach luminosities of $10^{52}$~erg~s$^{-1}$ and the mean neutrino energy of the order of 20~MeV. Under these conditions, XRFs and BdHNe become astrophysical laboratories for MeV-neutrino physics additional to core-collapse SNe.

On the other hand, the emission of TeV-PeV neutrinos is relevant for the observations of detectors such as the IceCube \citep{2013PhRvL.111b1103A}. High-energy neutrino emission mechanisms have been proposed within the context of the traditional model of long GRBs. In the traditional ``collapsar'' scenario \citep{1993ApJ...405..273W,Paczynski:1998ey,1999ApJ...524..262M} the gravitational collapse of a single, fast rotating, massive star originates a BH surrounded by a massive accretion disk \citep[see, e.g.,][for a review]{2004RvMP...76.1143P}, and the GRB dynamics follows the ``fireball'' model that assumes the existence of an ultra-relativistic collimated jet with Lorentz factor $\Gamma\sim 10^2$--$10^3$ \citep[see e.g.][]{1990ApJ...365L..55S,1993MNRAS.263..861P,1993ApJ...415..181M,1994ApJ...424L.131M}. This scenario has been adopted for the explanation of the prompt emission, as well as both the afterglow and the GeV emission of long GRBs. The GRB light-curve structures are there described by (internal or external) shocks \citep[see, e.g.,][]{1992MNRAS.258P..41R,1994ApJ...430L..93R}. The high-energy neutrinos in this context are produced from the interaction of shock-accelerated cosmic-rays (e.g. protons) with the interstellar medium \citep[see e.g.][and references therein]{2017APh....86...11A,2015PhR...561....1K}. A recent analysis of the thermal emission of the X-ray flares observed in the early afterglow of long GRBs (at source rest-frame times $t\sim 10^2$~s) show that it occurs at radii $\sim10^{12}$~cm and expands with a mildly-relativistic $\Gamma\lesssim 4$ \citep[see][for further details]{2017arXiv170403821R}. This rules out the ultra-relativistic expansion in the GRB afterglow traditionally adopted in the literature. Interestingly, the aforementioned mechanisms of high-energy neutrino production conceived in the collapsar-fireball model can still be relevant in the context of BdHNe and authentic short GRBs (S-GRBs, NS-NS mergers with $E_{\rm iso}\gtrsim 10^{52}$~erg leading to BH formation; see \citealp{2016ApJ...832..136R}, for the classification of long and short bursts in seven different sub-classes). The emission in the $0.1$--$100$~GeV energy band observed in these two GRB sub-classes has been shown to be well explained by a subsequent accretion process onto the newly-born BH (\citealp{2015ApJ...798...10R,2015ApJ...808..190R,2016ApJ...831..178R,2016ApJ...832..136R,2017ApJ...844...83A}; see also Aimuratov et al. in preparation). Such GeV emission is not causally connected either with the prompt emission or with the afterglow emission comprising the flaring activity \citep{2017arXiv170403821R}. An ultra-relativistic expanding component is therefore expected to occur in BdHNe and S-GRBs which deserves to be explored in forthcoming studies as a possible source of high-energy neutrinos. Specifically, this motivates the present article to identify the possible additional channels to be explored in the hypercritical accretion not around a NS but around a BH.

The aim of this article is to extend the analysis of the MeV-neutrino emission in the hypercritical accretion process around a NS in the XRFs and BdHNe to assess the possible occurrence of neutrino flavor oscillations.

We shall show in this work that, before escaping to the outer space, i.e. outside the Bondi-Hoyle accretion region, the neutrinos experience an interesting phenomenology. The neutrino density near the NS surface is so high that the neutrino self-interaction potential, usually negligible in other very well-known scenarios like the Sun, the upper layers of Earth's atmosphere and terrestrial reactor and accelerator experiments, becomes more relevant than the matter potential responsible for the famous Mikheyev-Smirnov-Wolfenstein (MSW) effect~\citep{Wolfenstein:1977ue,Mikheev:1986wj}. A number of papers have been dedicated to the consequences of the neutrino self-interaction dominance~\citep{Notzold:1987ik,1992PhLB..287..128P,Qian:1994wh,Pastor:2002we,Duan:2005cp,Sawyer:2005jk,Fuller:2005ae,Fogli:2007bk,Duan:2007fw,Raffelt:2007yz,EstebanPretel:2007ec,EstebanPretel:2007yq,Chakraborty:2008zp,Duan:2008eb,Duan:2007sh,Dasgupta:2008my,Dasgupta:2007ws,Sawyer:2008zs,Duan:2010bg,Wu:2011yi}, most of them focused on SN neutrinos. In these cases, the SN induces the appearance of collective effects such as synchronized and bipolar oscillations leading to an entirely new flavor content of emitted neutrinos when compared with the spectrum created deep inside the star. The density of neutrinos produced in the hypercritical accretion process of XRFs and BdHNe is such that the neutrino self-interactions, as in the case of SNe, dominate the neutrino flavor evolution, giving rise to the aforementioned collective effects. The main neutrino source, in this case, is the $\nu\bar\nu$ pair production from $e^{-}e^{+}$ annihilation \citep{2016ApJ...833..107B} which leads to an equal number of neutrinos and antineutrinos of each type. This equality does not happen in the SN standard scenario. We will show that bipolar oscillations, inducing very quick flavor pair conversions $\nu_e\bar{\nu}_e\leftrightarrow \nu_\mu\bar{\nu}_\mu\leftrightarrow \nu_\tau\bar{\nu}_\tau$, can occur with oscillation length as small as $O(0.05$--$1)$~kilometers. However, the $\nu$--$\bar\nu$ symmetry characterizing our system leads to the occurrence of kinematic decoherence making the neutrino flavor content to reach equipartition deep inside the accretion zone. In the regions far from the NS surface where the neutrino density is not so high, the matter potential turns to dominate and MSW resonances can take place. As a result, an entirely different neutrino flavor content emerges from the Bondi-Hoyle surface when compared with what was originally created in the bottom of the accretion zone. 

This article is organized as follows. In Sec.~\ref{sec:2} we outline the general features of the accretion process onto the NS within the IGC paradigm and present the processes responsible for the neutrino creation. From these features, we obtain the distribution functions that describe the neutrino spectrum near the NS surface. Sec.~\ref{sec:oscillations} shows a derivation of the equations that drive the evolution of neutrino oscillations closely related to the geometrical and physical characteristics of our system. We discuss some details on the neutrino oscillation phenomenology. Since we have to face a nonlinear integro-differential system of equations of motion, we introduce the single-angle approximation to later recover the full realistic phenomenology after generalizing our results to the multi-angle approach and, consequently, de-coherent picture. In Sec.~\ref{sec:spectra} the final neutrino emission spectra are presented and compared with those ones in which neutrinos are created in the accretion zone. Finally, we present in Sec.~\ref{sec:4} the conclusions and some perspectives for future research on this subject. 

\section{Neutrino creation during hypercritical accretion}\label{sec:2}

\begin{figure}
\centering
\includegraphics[width=\hsize,clip]{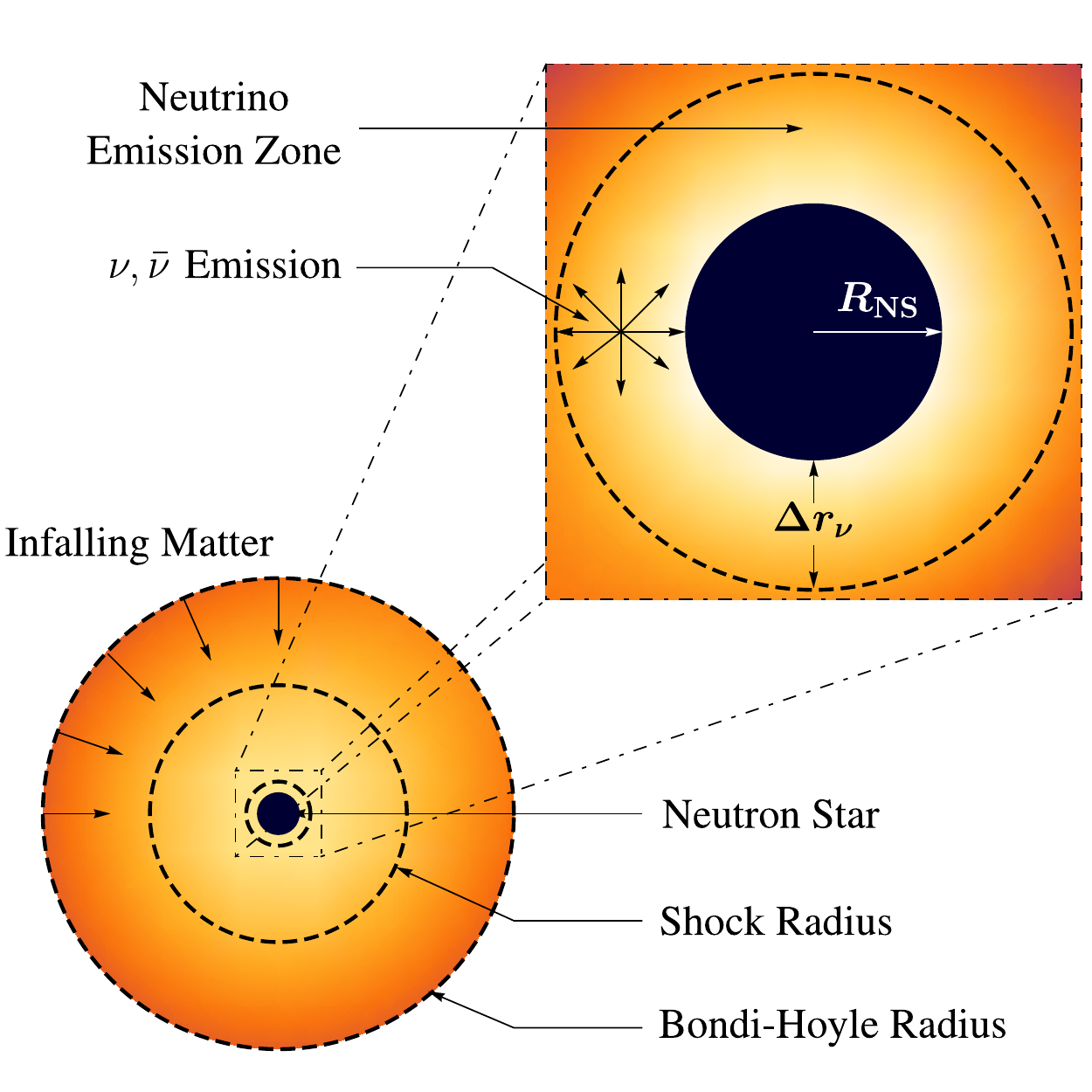}
\caption{Schematic representation of the accretion process onto the NS and the neutrino emission. The supernova ejected material reaches the NS Bondi-Hoyle radius and falls onto the NS surface. The material shocks and decelerates as it piles over the NS surface. At the neutrino emission zone, neutrinos take away most of the infalling matter's energy. The neutrino emission allows the material to reduce its entropy to be incorporated to the NS. The image is not to scale. For binary system with $M_{\rm NS}=2M_{\odot}$ and $R_{\rm NS}=10$ km, and a $M_{\rm ZAMS}=20M_{\odot}$ progenitor, at $\dot{M}=10^{-2}M_{\odot}/{\rm s}$, the position of the Bondi-Hoyle and Shock radii are $2.3\times 10^5$ km and $31$ km, respectively. The neutrino emission zone's thickness is $\Delta r_{\nu}=0.8$ km.}
\label{fig:IGC}
\end{figure}

The SN material first reaches the gravitational capture region of the NS companion, namely the Bondi-Hoyle region. The infalling material shocks as it piles up onto the NS surface forming an accretion zone where it compresses and eventually becomes sufficiently hot to trigger a highly efficient neutrino emission process. Neutrinos take away most of the infalling matter's gravitational energy gain, letting it reduce its entropy and be incorporated into the NS. Fig.~\ref{fig:IGC} shows a sketch of this entire hypercritical accretion process.

It was shown in \citet{2016ApJ...833..107B} that the matter in the accretion zone near the NS surface develops conditions of temperature and density such that it is in a non-degenerate, relativistic, hot plasma state. The most efficient neutrino emission channel under those conditions becomes the electron positron pair annihilation process:
\begin{equation}
e^{-}\!e^{+}\! \rightarrow \nu\,\bar{\nu}.
\label{process}
\end{equation}

The neutrino emissivity produced by this process is proportional to the accretion rate to the 9/4 power (see below). This implies that the higher the accretion rate the higher the neutrino flux, hence the largest neutrino flux occurs at the largest accretion rate.

We turn now to estimate the accretion rate and thus the neutrino emissivity we expect in our systems.

\subsection{Accretion rate in XRFs and BdHNe}\label{sec:2a}

We first discuss the amount of SN matter per unit time reaching the gravitational capture region of the NS companion, namely the Bondi-Hoyle accretion rate. It has been shown in \citet{2015ApJ...805...98B,2016ApJ...833..107B} that the shorter (smaller) the orbital period (separation) the higher the peak accretion rate $\dot{M}_{\rm peak}$ and the shorter the time at which it peaks, $t_{\rm peak}$.

The Bondi-Hoyle accretion rate is proportional to the density of the accreted matter and inversely proportional to its velocity. Thus, we expect the accretion rate to increase as the denser and slower inner layers of the SN reach the accretion region. Based on these arguments, \citet{2016ApJ...833..107B} derived simple, analytic formulas for $\dot{M}_{\rm peak}$ and $t_{\rm peak}$ as a function of the orbital period (given all the other binary parameters) that catch both the qualitatively and quantitatively behaviors of these two quantities obtained from full numerical integration. We refer the reader to the Appendix A of that article for further details. For the scope of this work these analytic expressions are sufficient to give us an estimate of the hypercritical accretion rates and related time scale developed in these systems:
\begin{subequations}
\begin{gather}
t_{\rm peak} \approx \left(1-\frac{2 M_{\rm NS}}{M}\right)\left(\frac{G M}{4\pi^2}\right)^{1/3} \left(\frac{R_{\rm star}^0}{\eta R_{\rm core}}\right) \frac{P^{2/3}}{v_{\rm star,0}},\\
\dot{M}_{\rm peak} \approx  2\pi^2\frac{(2 M_{\rm NS}/M)^{5/2}}{(1-2 M_{\rm NS}/M)^3}\eta^{3-m}\frac{\rho_{\rm core}\,R_{\rm core}^3}{P},
\end{gather}\label{eq:mpeakandtpeak}
\end{subequations}
where $P$ is the orbital period, $m$ is the index of the power-law density profile of the pre-SN envelope,  $v_{\rm star,0}$ is the velocity of the outermost layer of the SN ejecta, $M=M_{\rm CO}+M_{\rm NS}$ is the total binary mass, $M_{\rm CO} = M_{\rm env} + M_{\nu\rm NS}$ is the total mass of the CO$_{\rm core}$ given by the envelope mass and the mass of the central remnant, i.e. the new NS (hereafter $\nu$NS) formed from the region of the CO$_{\rm core}$ which undergoes core-collapse (i.e. roughly speaking the iron core of density $\rho_{\rm core}$ and radius $R_{\rm core}$). We here adopt $ M_{\nu\rm NS} = 1.5~M_\odot$. The parameter $\eta$ is given by 
\begin{equation}\label{eq:rinner}
\eta \equiv \frac{R_{\rm star}^0}{R_{\rm core}}\frac{1+m}{1+m (R_{\rm star}^0/\hat{R}_{\rm core})},
\end{equation}
where $R_{\rm star}^0$ is the total radius of the pre-SN CO$_{\rm core}$; $\hat{\rho}_{\rm core}$ and $\hat{R}_{\rm core}$ are parameters of the pre-SN density profile introduced to account of the finite size of the envelope, and $m$ is the power-law index followed by the density profile at radii $r>R_{\rm core}$ \citep[see][for further details]{2016ApJ...833..107B}.

\begin{figure}
\centering	
\includegraphics[width=\hsize,clip]{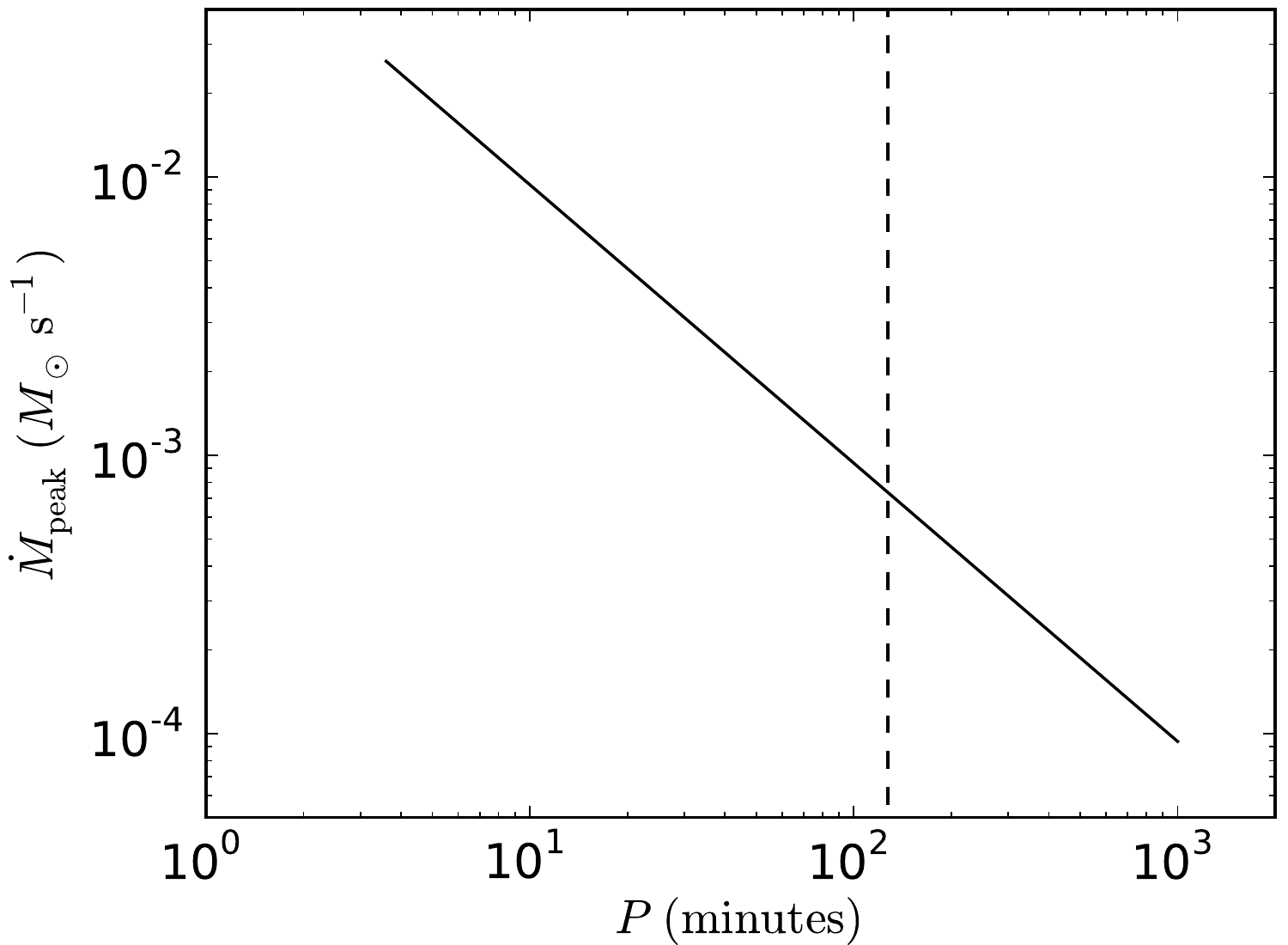}
\caption{Peak accretion rate, $\dot{M}_{\rm peak}$, as a function of the binary orbital period, as given by  Eq.~(\ref{eq:mpeakandtpeak}). This example corresponds to the following binary parameters: a CO$_{\rm core}$ formed by a $M_{\rm ZAMS}=20~M_\odot$ progenitor, i.e. $M_{\rm CO}=5.4~M_\odot$, an initial NS mass $2.0~M_\odot$, $v_{\rm star,0}=2\times10^9$~cm~s$^{-1}$, $\eta\approx 0.41$ and index $m=2.946$ \citep[see][for further details]{2016ApJ...833..107B}. For these parameters the largest orbital period for the induced collapse of the NS to a BH by accretion is $P_{\rm max}\approx 127$~min which is represented by the vertical dashed line.}
\label{fig:Mpeak}
\end{figure}

Fig.~\ref{fig:Mpeak} shows the peak accretion rate in Eq.~(\ref{eq:mpeakandtpeak}) as a function of the orbital period. In this example, we consider the following binary parameters \citep[see][for details]{2016ApJ...833..107B}: a CO$_{\rm core}$ produced by a zero-age main-sequence (ZAMS) progenitor with $M_{\rm ZAMS}=20~M_\odot$, i.e. $M_{\rm CO}=5.4~M_\odot$, an initial NS mass $2.0~M_\odot$, and a velocity of the outermost ejecta layer $v_{\rm star,0}=2\times10^9$~cm~s$^{-1}$. For these parameters, $\eta\approx 0.41$.

It was shown in \citet{2015ApJ...812..100B,2016ApJ...833..107B} the existence of a maximum orbital period, $P_{\rm max}$, over which the accretion onto NS companion is not high enough to bring it to the critical mass for gravitational collapse to a BH. As we have recalled in the Introduction, CO$_{\rm core}$-NS binaries with $P > P_{\rm max}$ lead to XRFs while the ones with $P \lesssim P_{\rm max}$ lead to BdHNe. For the binary parameters of the example in Fig.~\ref{fig:Mpeak}, $P_{\rm max}\approx 127$~min (vertical dashed line). We can therefore conclude that BdHNe can have peak accretion rates in the range $\dot{M}_{\rm peak}\sim 10^{-3}$-- few~$10^{-2}~M_\odot$~s$^{-1}$ while XRFs would have $\dot{M}_{\rm peak}\sim 10^{-4}$--$10^{-3}~M_\odot$~s$^{-1}$.

\subsection{Neutrino emission at maximum accretion}\label{sec:2b}

For the accretion rate conditions characteristic of our models at peak $\sim 10^{-4}$--$10^{-2}~M_\odot$~s$^{-1}$, pair annihilation dominates the neutrino emission and electron neutrinos remove the bulk of the energy \citep{2016ApJ...833..107B}. The $e^+e^-$ pairs producing the neutrinos are thermalized at the matter temperature. This temperature is approximately given by:
\begin{equation} \label{eq:Tacc}
T_{\rm acc}\approx \left(\frac{3 P_{\rm shock}}{4 \sigma/c}\right)^{1/4}=\left(\frac{7}{8} \frac{\dot{M}_{\rm acc}
v_{\rm acc} c}{4 \pi R^2_{\rm NS} \sigma}\right)^{1/4},
\end{equation}
where $P_{\rm shock}$ is the pressure of the shock developed on the accretion zone above the NS surface, $\dot{M}_{\rm acc}$ is the accretion rate, $v_{\rm acc}$ is the velocity of the infalling material, $\sigma$ is the Stefan-Boltzmann constant and $c$ the speed of light. It can be checked that, for the above accretion rates, the system develops temperatures and densities ($T\gtrsim 10^{10}$~K and $\rho\gtrsim 10^6$~g~cm$^{-3}$; see e.g. Fig. 16 in \citealp{2016ApJ...833..107B}) for which the neutrino emissivity of the $e^+e^-$ annhiliation process can be estimated by the simple formula \citep{2001PhR...354....1Y}:
\begin{equation}\label{eq:L_neutrinos}
\epsilon_{e^{-}\!e^{+}} \approx 8.69\times 10^{30}\left(\frac{k_B T}{1\,{\rm MeV}}\right)^9\,\, {\rm MeV}\,{\rm cm}^{-3}\,{\rm s}^{-1},
\end{equation}
where $k_B$ is the Boltzmann constant.

The accretion zone is characterized by a temperature gradient with a typical scale height $\Delta r_{\rm ER} = T/\nabla T \approx 0.7~R_{\rm NS}$. Owing to the strong dependence of the neutrino emission on temperature, most of the neutrinos are emitted from a spherical shell around the NS of thickness (see Fig.~\ref{fig:IGC})
\begin{equation}
\Delta r_{\nu} = \frac{\epsilon_{e^{-}\!e^{+}}}{\nabla \epsilon_{e^{-}\!e^{+}}} = \frac{\Delta r_{\rm ER}}{9} \approx 0.08R_{\rm NS}.
\label{neutrinoshell}
\end{equation}
Eqs.~(\ref{eq:Tacc}) and (\ref{eq:L_neutrinos}) imply the neutrino emissivity satisfies $\epsilon_{e^{-}\!e^{+}} \propto \dot{M}^{9/4}_{\rm acc}$ as we had anticipated. These conditions lead to the neutrinos to be efficient in balancing the gravitational potential energy gain, allowing the hypercritical accretion rates. The effective accretion onto the NS can be estimated as 
\begin{equation}\label{eq:Mdoteff}
\dot{M}_{\rm eff} \approx \Delta M_{\nu} \frac{L_{\nu}}{E_g},
\end{equation}
where $\Delta M_{\nu}$, $L_{\nu}$ are, respectively, the mass and neutrino luminosity in the emission region, and $E_g=(1/2) G M_{\rm NS} \Delta M_{\nu}/(R_{\nu}+\Delta r_{\nu})$ is half the gravitational potential energy gained by the material falling from infinity to the $R_{\rm NS}+\Delta r_{\nu}$. The neutrino luminosity is
\begin{equation}
L_{\nu}\approx 4\pi R_{\rm NS}^2\Delta r_{\nu} \epsilon_{e^{-}\!e^{+}}.
\label{eq:neutrinoluminosity}
\end{equation}
with $\epsilon_{e^{-}\!e^{+}}$ being the neutrino emissivity in Eq.~(\ref{eq:L_neutrinos}). For $M_{\rm NS}=2~M_\odot$ and temperatures $1$--10~MeV, the Eqs.~(\ref{eq:Mdoteff}) and (\ref{eq:neutrinoluminosity}) result $\dot{M}_{\rm eff} \approx 10^{-10}$--$10^{-1}~M_\odot$~s$^{-1}$ and $L_{\nu} \approx 10^{48}$--$10^{57}$~MeV~s$^{-1}$.

\begin{table*}
\centering
\begin{tabular}{c c c c c c c c c c c c c c c}
\hline
  $\dot{M}$\T\B & $\rho$\T\B & $k_{B}T$\T\B & $\eta_{e^{\mp}}$\T\B & $n_{e^{-}}\!-n_{e^{+}}$\T\B & $k_{B}T_{\nu\bar{\nu}}$\T\B & $\langle E_\nu \rangle$\T\B & $F^{C}_{\nu_e,\bar{\nu}_e}$\T\B & $F^{C}_{\nu_x,\bar{\nu}_x}$\T\B & $n^{C}_{\nu_{e}\bar{\nu}_{e}}$\T\B & $n^{C}_{\nu_{x}\bar{\nu}_{x}}$\T\B & $\sum_{i}\,n^{C}_{\nu_{i}\bar{\nu}_{i}}$\T\B \\ 
	$(M_\odot$~s$^{-1}$)\B & (g~cm$^{-3})$\B & (MeV)\B & \B & (cm$^{-3}$)\B & (MeV)\B & (MeV)\B & (cm$^{-2}$s$^{-1}$)\B & (cm$^{-2}$s$^{-1}$)\B & (cm$^{-3})$\B & (cm$^{-3})$\B & (cm$^{-3})$\B \\ \hline\hline
	$10^{-8}$\T & $1.46\times10^{6}$\T & 1.56\T & $\mp 0.325$\T & $4.41\times10^{29}$\T & 1.78\T & 6.39\T & $4.17\times 10^{36}$\T & $1.79\times 10^{36}$\T& $2.78\times10^{26}$\T & $1.19\times10^{26}$\T & $3.97\times10^{26}$\T \\
   $10^{-7}$ & $3.90\times10^{6}$ & 2.01 & $\mp 0.251$ & $1.25\times10^{30}$ & 2.28 & 8.24 & $3.16\times 10^{37}$ & $1.36\times 10^{37}$ & $2.11\times10^{27}$ & $9.00\times10^{26}$ & $3.01\times10^{27}$ \\
    $10^{-6}$ & $1.12\times10^{7}$ & 2.59 & $\mp 0.193$ & $3.38\times10^{30}$ & 2.93 & 10.61 & $2.40\times 10^{38}$ & $1.03\times 10^{38}$ & $1.60\times10^{28}$ & $6.90\times10^{27}$ & $2.29\times10^{28}$ \\
    $10^{-5}$ & $3.10\times10^{7}$ & 3.34 & $\mp 0.147$ & $9.56\times10^{30}$ & 3.78 & 13.69 & $1.84\times 10^{39}$ & $7.87\times 10^{38}$ & $1.23\times10^{29}$ & $5.20\times10^{28}$ & $1.75\times10^{29}$ \\
    $10^{-4}$ & $8.66\times10^{7}$ & 4.30 & $\mp 0.111$ & $2.61\times10^{31}$ & 4.87 & 17.62 &  $1.39\times 10^{40}$ & $5.94\times 10^{39}$ & $9.24\times10^{29}$ & $3.96\times10^{29}$ & $1.32\times10^{30}$ \\
    $10^{-3}$ & $2.48\times10^{8}$ & 5.54 & $\mp 0.082$ & $7.65\times10^{31}$ & 6.28 & 22.70 & $1.04\times 10^{41}$ & $4.51\times 10^{40}$ & $7.00\times10^{30}$ & $3.00\times10^{30}$ & $1.00\times10^{31}$ \\
  $10^{-2}$ & $7.54\times10^{8}$ & 7.13 & $\mp 0.057$ & $2.27\times10^{32}$ & 8.08 & 29.22 & $7.92\times 10^{41}$  & $3.39\times 10^{41}$ & $5.28\times10^{31}$ & $2.26\times10^{31}$ & $7.54\times10^{31}$ \\ \hline
\end{tabular}
\caption{Characteristics inside the neutrino emission zone and the neutrino spectrum for selected values of the accretion rate $\dot M$. The electron fraction is $Y_{e}=0.5$, the pinching parameter for the neutrino spectrum is $\eta_{\nu\bar{\nu}}=2.0376$ and the.}
\label{tab:tab1}
\end{table*}

\subsection{Neutrino spectrum at the NS surface}\label{sec:2c}

After discussing the general features of neutrino emission during the accretion process, it is necessary for our analysis of the neutrino oscillations to determine the neutrino spectrum at the NS surface. Specifically, we need to determine the ratios at which the neutrinos of each flavor are created and their average energy so that we can find a fitting distribution function $f_{\nu}$ with these characteristics. 

Since the main source of neutrinos is the $e^{-}\!e^{+}$ pair annihilation process we can conclude that neutrinos and antineutrinos are created in equal number. Furthermore, the information about the neutrino and antineutrino emission of a given flavor $i$ can be calculated from the integral \citep{2001PhR...354....1Y}:
\begin{equation}
\varepsilon^{m}_{i} = \frac{2G^{2}_{F}\left(m_{e}c^{2}\right)^{4}}{3\left(2\pi\hbar\right)^{7} \left(\hbar c\right)^{3}} \int\!\! f_{e^{-}}f_{e^{+}}\frac{\left( E^{m}_{e^{-}} + E^{m}_{e^{+}} \right)}{E_{e^{-}}E_{e^{+}}}\sigma_{i}\,d^{3}\mathbf{p}_{e^{-}}d^{3}\mathbf{p}_{e^{+}}
\label{eq:emissionratenu}
\end{equation}
where $G_{F} = 8.963\times10^{-44}$~MeV~cm$^{3}$ is the Fermi constant of weak interactions. Here $m=0,1,\ldots$ and should not be confused with the index of the power-law density profile of the pre-SN envelope in Sec. \ref{sec:2a}). $f_{e^{\pm}}$ are the Fermi-Dirac distributions for electron and positrons
\begin{equation}
f_{e^{\mp}} = \frac{1}{1+\exp\left(\frac{E_{e^{\mp}}}{k_{B}T}\mp\eta_{e^{\mp}}\right)}.
\label{eq:FermiDiracDistribution}
\end{equation}

$\eta_{e^{\mp}}$ is the electron (positron) degeneracy parameter including it's rest mass. The Dicus cross section $\sigma_{i}$ is written in terms of the electron and positron four-momenta $p_{e^{\pm}} = \left(E_{e^{\pm}}/c,\mathbf{p}_{e^{\pm}}\right)$ as \citep{Dicus:1972yr}
\begin{align}
\sigma_{i} = C^{2}_{+,i}\left(1+3\frac{p_{e^{-}}\cdot p_{e^{+}}}{\left(cm_{e}\right)^{2}}\right.&\left. +2\frac{\left(p_{e^{-}}\cdot p_{e^{+}}\right)^{2}}{\left(cm_{e}\right)^{4}}\right)\nonumber\\
&+3C^{2}_{-,i}\left(1+\frac{p_{e^{-}}\cdot p_{e^{+}}}{\left(cm_{e}\right)^{2}}\right).\label{eq:Dcrosssection}\end{align}

The factors $C^{2}_{\pm,i}$, are written in terms of the weak interaction vector and axial-vector constants: $C^{2}_{\pm,i}=C^{2}_{V_{i}} \pm C^{2}_{A_{i}}$, where $C_{V_{e}}\! = 2\sin^2\theta_{\rm W} + 1/2$, $C_{A_{e}}\! = 1/2$, $C_{V_{\mu}}\! = C_{V_{\tau}}\! = C_{V_{e}} - 1$ and $C_{A_{\mu}}\! = C_{A_{\tau}}\! = C_{A_{e}} - 1$ with the numerical value of the Weinberg angle approximated by $\sin^2\theta_{\rm W} \approx 0.231$ \citep{Olive:2016xmw}.

For $m=0$ and $m=1$ Eq. (\ref{eq:emissionratenu}) gives the neutrino and antineutrino number emissivity (neutrino production rate), and the neutrino and antineutrino energy emissivity (energy per unit volume per unit time) for a certain flavor $i$, respectively. Hence, not only we are able to calculate the total number emissivity with
\begin{equation}
n=\!\!\sum_{i\in\{e,\tau,\mu\}}\!\varepsilon^{0}_{i},
\label{eq:toalnumberemissivity}
\end{equation}
but we can also calculate the neutrino or antineutrino energy moments with
\begin{equation}
\langle E^{m}_{\nu_{i}\left(\bar{\nu}_{i}\right)} \rangle = \frac{\varepsilon^{m}_{i}}{\varepsilon^{0}_{i}},\,\, {\rm for }\,\, m\geq 1.
\label{eq:neutrinomoments}
\end{equation}

We wish to construct a Fermi-Dirac like fitting formula for the neutrino spectrum as it is usually done in supernovae neutrino emission \citep{1989A&AS...78..375J,1989A&A...224...49J}. That is, a function like Eq.~(\ref{eq:FermiDiracDistribution}) in terms of two parameters: the effective neutrino temperature $T_{\nu\bar{\nu}}$ and the effective neutrino degeneracy parameter $\eta_{\nu\bar{\nu}}$ otherwise known as the \emph{pinching} parameter \citep{Raffelt:1996wa,Keil:2002in}. To that end, it is enough to calculate the first two moments. In particular, for a relativistic non-degenerate plasma ($k_{B}T>2m_{e}c^2$ and $1>\eta_{e^{\mp}}$ see table \ref{tab:tab1}) Eq. (\ref{eq:emissionratenu}) can be approximated with a very good accuracy by \citep{2001PhR...354....1Y}
\begin{align}
\varepsilon^{m}_{i} \approx \frac{2G^{2}_{F}\left(k_{B}T\right)^{8+m}}{9\pi^{5}\hbar\left(\hbar c\right)^{9}}C^{2}_{+,i}&\left[\mathcal{F}_{m+1}\left(\eta_{e^{+}}\right)\mathcal{F}_{1}\left(\eta_{e^{-}}\right)\right.\nonumber\\
&\left.+\mathcal{F}_{m+1}\left(\eta_{e^{-}}\right)\mathcal{F}_{1}\left(\eta_{e^{+}}\right)\right]\label{eq:approximationyakovlev}
\end{align}
where $\mathcal{F}_{k}\left( \eta \right) = \int_0^\infty\!\!dx \, x^{k} / \left[1+\exp\left(x-\eta\right)\right]$ are the Fermi-Dirac integrals. For $m=1$, $\eta_{e^{\pm}}=0$ and adding over every flavor this expression reduces to Eq.~(\ref{eq:L_neutrinos}). With Eqs.~(\ref{eq:neutrinomoments}) and (\ref{eq:approximationyakovlev}) we find
\begin{subequations}
\begin{gather}
\langle E_{\nu}  \rangle = \langle E_{\bar{\nu}}  \rangle \approx 4.1\,k_{B}T\\
\langle E^{2}_{\nu} \rangle =  \langle E^{2}_{\bar{\nu}}  \rangle \approx 20.8\left( k_{B}T \right)^{2},
\end{gather}\label{eq:neutrinotwomoments}\end{subequations}
regardless of the neutrino flavor. Furthermore, we can calculate the ratio of emission rates between electronic and nonelectronic neutrino flavors in terms of the weak interaction constants
\begin{equation}
\frac{\varepsilon^{0}_{e}}{\varepsilon^{0}_{x}} = \frac{\varepsilon^{0}_{e}}{\varepsilon^{0}_{\mu}+\varepsilon^{0}_{\tau}} = \frac{C^{2}_{+,e}}{C^{2}_{+,\mu} + C^{2}_{+,\tau}} \approx \frac{7}{3}.
\label{eq:neutrinoratio}
\end{equation}

Some comments must be made about the results we have obtained:

\begin{itemize}

\item It is well known that, within the Standard Model of Particles, there are three neutrino flavors $\nu_{e},\bar{\nu}_{e}$, $\nu_{\mu},\bar{\nu}_{\mu}$ and $\nu_{\tau},\bar{\nu}_{\tau}$. However, as in Eq. (\ref{eq:neutrinoratio}), we will simplify our description using only two flavors: the electronic neutrinos and antineutrinos $\nu_{e},\bar{\nu}_{e}$, and a superposition of the other flavors $\nu_{x},\bar{\nu}_{x}$ $(x=\mu+\tau)$. This can be understood as follows. Since the matter in the accretion zone is composed by protons, neutrons, electrons and positrons, $\nu_e$ and $\bar\nu_e$ interact with matter by both charged and neutral currents, while $\nu_\mu$, $\nu_\tau$, $\bar\nu_\mu$ and $\bar\nu_\tau$ interact only by neutral currents. Therefore, the behavior of these states can be clearly divided into electronic and non-electronic. This distinction will come in handy when studying neutrino oscillations.

\item Representing the neutrino (antineutrino) density and flux in the moment of their creation with $n^{c}_{\nu_{i}(\bar{{\nu_{i}}})}$ and $F^{c}_{\nu_{i}(\bar{\nu_{i}})}$ respectively and using Eq.~(\ref{eq:neutrinoratio}) we can recollect two important facts:
\begin{subequations}
\begin{gather}
n^{C}_{\nu_{i}}=n^{C}_{\bar{{\nu_{i}}}}, \;\, F^{C}_{\nu_{i}} = F^{C}_{\bar{\nu_{i}}} \;\,\, \forall i\, \in \{ e,\mu,\tau \} \\
\frac{n^{C}_{\nu_e}}{n^{C}_{\nu_x}} =\frac{n^{C}_{\bar{\nu}_e}}{n^{C}_{\bar{\nu}_x}} = \frac{F^{C}_{\nu_e}}{F^{C}_{\nu_x}} = \frac{F^{C}_{\bar{\nu}_e}}{F^{C}_{\bar{\nu}_x}} \approx \frac{7}{3}.
\end{gather}\label{eq:creationdensityflux}\end{subequations}
Eqs.~(\ref{eq:creationdensityflux}) imply that, in the specific environment of our system, of the total number of neutrinos+antineutrinos emitted, $N_{\nu} + N_{\bar\nu}$, 70\% are electronic neutrinos ($N_{\nu_e}+N_{\bar\nu_e}$), 30\% are non-electronic ($N_{\nu_x}+N_{\bar\nu_x}$), while the total number of neutrinos is equal to the total number of antineutrinos, i.e. $N_{\nu}=N_{\bar\nu}$, where $N_{\nu} = N_{\nu_e}+N_{\nu_x}$ and $N_{\bar\nu} = N_{\bar\nu_e}+N_{\bar\nu_x}$.

\item Bearing in mind such high neutrino energies as the ones suggested by Eqs.~(\ref{eq:neutrinotwomoments}) , from here on out we will use the approximation
\begin{equation}
E_{\nu} \approx c\vert \mathbf{p} \vert \gg m_{\nu}c^{2},
\end{equation}
where $\mathbf{p}$ is the neutrino momentum.

\item From Eq.~(\ref{eq:neutrinomoments}) we obtain the same energy moments for both neutrinos and antineutrinos but, as \cite{PhysRevD.74.043006} points out, these energies should be different since, in reality, this expression returns the arithmetic mean of the particle and antiparticle energy moments, that is $\left( \langle E^{m}_{\nu} \rangle + \langle E^{m}_{\bar{\nu}} \rangle \right)/2$. However, if we calculate the differences between the energy moments with equations (41) and (46) in \cite{PhysRevD.74.043006} for the values of $T$ and $\eta_{e^{\pm}}$ we are considering, we get $\Delta\langle E \rangle \sim 10^{-2}$--$10^{-3}$~MeV and $\Delta\langle E^{2} \rangle \sim 10^{-3}$--$10^{-4}$~MeV$^{2}$. These differences are small enough that we can use the same effective temperature and pinching parameter for both neutrinos and antineutrinos.

\end{itemize}

Solving the equations
\begin{subequations}
\begin{gather}
4.1k_{B}T = k_{B}T_{\nu\bar{\nu}}\frac{\mathcal{F}_{3}\left(\eta_{\nu\bar{\nu}}\right)}{\mathcal{F}_{2}\left(\eta_{\nu\bar{\nu}}\right)}\\
20.8\left(k_{B}T\right)^{2} = \left(k_{B}T_{\nu\bar{\nu}}\right)^{2} \frac{\mathcal{F}_{4}\left(\eta_{\nu\bar{\nu}}\right)}{\mathcal{F}_{2}\left(\eta_{\nu\bar{\nu}}\right)}
\end{gather}\label{eq:eqsystem}\end{subequations}
for any value of $T$ in table (\ref{tab:tab1}) we find $T_{\nu\bar{\nu}} = 1.1331T$ and $\eta_{\nu\bar{\nu}}= 2.0376$. Integrating Eq.~(\ref{eq:FermiDiracDistribution}) over the neutrino momentum space using these values should give the neutrino number density. To achieve this we normalize it with the factor $1/\left(2\pi^{2}\left(k_{B}T_{\nu\bar{\nu}}\right)^{3}\mathcal{F}_{2}\left(\eta_{\nu\bar{\nu}}\right)\right)$ and then we multiply by
\begin{equation}
n^{C}_{\nu_{i}\left(\bar{\nu}_{i}\right)} = w_{\nu_{i}\left(\bar{\nu}_{i}\right)}\frac{L_{\nu}}{4\pi R^{2}_{\rm NS}\langle E_{\nu} \rangle\langle v \rangle} = w_{\nu_{i}\left(\bar{\nu}_{i}\right)}\frac{\varepsilon^{0}_{i}\Delta r_{\nu}}{c/2},
\label{eq:normalizedneutrinospectrum}
\end{equation}
where the neutrino's average radial velocity at $r=R_{\rm NS}$ is $\langle v \rangle = c/2$ \citep{Dasgupta:2008cu} and $w_{\nu_{e}}=w_{\bar{\nu}_{e}}=0.35$ and $w_{\nu_{x}}=w_{\bar{\nu}_{x}}=0.15$. To calculate the neutrino fluxes we simply $F^{C}_{\nu\left(\bar{\nu}_{i}\right)}=\langle v \rangle n^{c}_{\nu_{i}\left(\bar{\nu}_{i}\right)}$. Gathering our results we can finally write the distribution functions as
\begin{subequations}
\begin{gather}
f_{\nu_{e}} = f_{\bar{\nu}_{e}} = \frac{2\pi^{2}\left(\hbar c\right)^{3}n^{C}_{\nu_{e}}}{\left(k_{B}T_{\nu\bar{\nu}}\right)^{3}\mathcal{F}_{2}\left(\eta_{\nu\bar{\nu}}\right)}\frac{1}{1+\exp\left(E/k_{B}T_{\nu\bar{\nu}}-\eta_{\nu\bar{\nu}}\right)}\\
f_{\nu_{x}} = f_{\bar{\nu}_{x}} = \frac{2\pi^{2}\left(\hbar c\right)^{3}n^{C}_{\nu_{x}}}{\left(k_{B}T_{\nu\bar{\nu}}\right)^{3}\mathcal{F}_{2}\left(\eta_{\nu\bar{\nu}}\right)}\frac{1}{1+\exp\left(E/k_{B}T_{\nu\bar{\nu}}-\eta_{\nu\bar{\nu}}\right)}\end{gather}\label{eq:neutrinofullinitialspectrum}\end{subequations}

It can be checked that these distributions obey
\begin{subequations}
\begin{gather}
\int\! f_{\nu_{i}} \frac{d^{3}\mathbf{p}}{\left(2\pi\hbar\right)^{3}} = n^{C}_{\nu_{i}}\\
\int\! E f_{\nu_{i}} \frac{d^{3}\mathbf{p}}{\left(2\pi\hbar\right)^{3}} = \langle E_{\nu} \rangle n^{C}_{\nu_{i}} = \varepsilon^{1}_{i}\label{eq:distributionconditions}\end{gather}\end{subequations}
and with these conditions satisfied we can conclude that Eqs.~(\ref{eq:neutrinofullinitialspectrum}) are precisely the ones that emulate the neutrino spectrum at the NS surface. In Table \ref{tab:tab1} we have collected the values of every important quantity used in the calculations within this section for the range of accretion rates in which we are interested. 

Considering that the problem we attacked in this section reduces to finding a normalized distribution whose first two moments are fixed, the choice we have made with Eqs.~(\ref{eq:neutrinofullinitialspectrum}) is not unique. The solution depends on how many moments are used to fit the distribution and what kind of function is used as an ansatz. A different solution based on a Maxwell-Boltzmann distribution can be found in \cite{Keil:2002in,Fogli:2004ff,PhysRevD.74.043006}.

At this stage, we can identify two main differences between neutrino emission in SNe and in the IGC process of XRFs and BdHNe, within the context of neutrino oscillations. The significance of these differences will become clearer in next sections but we mention them here to establish a point of comparison between the two systems since SN neutrino oscillations have been extensively studied.

\begin{itemize}
\item Neutrinos of all flavors in XRFs and BdHNe have the same temperature, which leads to equal average energy. The neutrinos produced in SNe are trapped and kept in thermal equilibrium within their respective neutrino-sphere. The neutrino-spheres have different radii, causing different flavors to have different average energies. This energy difference leads to a phenomenon called \emph{spectral stepwise swap} which, as we will show below, is not present in our systems \citep[see, e.g.,][and references therein]{Raffelt:1996wa,Fogli:2007bk,Dasgupta:2007ws}.

\item As we have discussed above, in XRFs and BdHNe neutrinos and antineutrinos are emitted in equal number. Due to this fact, kinematical decoherence occurs (up to a number difference of 30\% this statement is valid; see Sec.~\ref{sec:solutions} for further details). Instead, SN neutrino and antineutrino fluxes differ such that $F_{\nu_{e}} > F_{\bar{\nu}_{e}} > F_{\nu_{x}} = F_{\bar{\nu}_{x}}$. It has been argued that this difference between neutrinos and antineutrinos is enough to dampen kinematical decoherence, so that bipolar oscillations are a feature present in SN neutrinos \citep[see, e.g.,][]{EstebanPretel:2007ec}.
\end{itemize}

In the next section, we will use the results presented here to determine the neutrino flavor evolution in the accretion zone. 

\section{Neutrino Oscillations}\label{sec:oscillations}

In recent years the picture of neutrino oscillations in dense media, based only on MSW effects, has undergone a change of paradigm by the insight that the refractive effects of neutrinos on themselves due to the neutrino self-interaction potential are crucial~\citep{Notzold:1987ik,1992PhLB..287..128P,Qian:1994wh,Pastor:2002we,Duan:2005cp,Sawyer:2005jk,Fuller:2005ae,Fogli:2007bk,Duan:2007fw,Raffelt:2007yz,EstebanPretel:2007ec,EstebanPretel:2007yq,Chakraborty:2008zp,Duan:2008eb,Duan:2007sh,Dasgupta:2008my,Dasgupta:2007ws,Sawyer:2008zs,Duan:2010bg,Wu:2011yi}.

As we discussed in Sec.~\ref{sec:2}, in our physical system of interest neutrinos are mainly created by electron-positron pair annihilation and so the number of neutrinos is equal to the number of antineutrinos. Such a fact creates an interesting and unique physical situation, different from, for example, SN neutrinos for which traditional models predict a predominance of electron neutrinos mainly due to the deleptonization caused by the URCA process \citep[see, e.g.,][]{EstebanPretel:2007ec}.

The neutrino self-interaction potential decays with the radial distance from the neutron star faster than the matter potential. This is a direct consequence of the usual $1/r^2$ flux dilution and the collinearity effects due to the neutrino velocity dependence of the potential. Consequently, we identify three different regions along the neutrino trajectory in which the oscillations are dominated by intrinsically different neutrino phenomenology. Fig.~\ref{fig:potentials} illustrates the typical situation of the physical system we are analyzing. Just after the neutrino creation in the regions of the accretion zone very close to the surface of the NS, neutrinos undergo kinematic decoherence along the same length scale of a single cycle of the so-called bipolar oscillations. Bipolar oscillations imply very fast flavor conversion between neutrino pairs $\nu_e\bar{\nu}_e\leftrightarrow \nu_\mu\bar{\nu}_\mu\leftrightarrow \nu_\tau\bar{\nu}_\tau$ and, amazingly, the oscillation length in this region can be so small as of the order tens of meters. 
Note that kinematic decoherence is just the averaging over flavor neutrino states process resulting from quick flavor conversion which oscillation length depends on the neutrino energy. It does not imply quantum decoherence and, thus, neutrinos are yet able to quantum oscillate if appropriate conditions are satisfied. In fact, as it can be observed from Figs.~\ref{fig:singleangle} and \ref{fig:singleanglesolutions} below, bipolar oscillations preserve the characteristic oscillation pattern, differently from quantum decoherence which would lead to a monotonous dumping figure.  

Kinematic decoherence is relevant when three conditions are met: (i) The self-interaction potential dominates over the vacuum potential. (ii) The matter potential does not fulfill the MSW condition. (iii) There is a low asymmetry between the neutrino and antineutrino fluxes. We will see that our system satisfies all three conditions.

As the self-interaction potential becomes small and the matter potential becomes important, oscillations are suppressed and we do not expect significant changes in the neutrino flavor content along this region. This situation changes radically when the matter potential is so small that it is comparable with neutrino vacuum frequencies $\Delta m^2/2p$, where $\Delta m^2$ is the neutrino squared mass difference and $p$ is the norm of the neutrino momentum $\mathbf{p}$. In this region, the neutrino self-interaction potential is negligible and the usual MSW resonances can occur. Therefore, we can expect a change in the neutrino spectrum.

We dedicate this section to a detailed derivation of the equations of motion (EoM) of flavor evolution. In later sections, we will analyze the neutrino oscillation phenomenology to build the neutrino emission spectrum from a binary hyper-accretion system.    

\subsection{Equations of motion}\label{sec:3a}

The equations of motion (EoM) that govern the evolution of an ensemble of mixed neutrinos are the quantum Liouville equations
\begin{subequations}
\begin{gather}
i\dot{\rho}_{\mathbf{p}} = [H_{\mathbf{p}},\rho_{\mathbf{p}}]\\
i\dot{\bar{\rho}}_{\mathbf{p}} = [\bar{H}_{\mathbf{p}},\bar{\rho}_{\mathbf{p}}]\end{gather}\label{eq:Liouville}\end{subequations}
where we have adopted the natural units $c=\hbar=1$. In these equations $\rho_{\mathbf{p}}$ ($\bar{\rho}_{\mathbf{p}}$) is the matrix of occupation numbers $(\rho_{\mathbf{p}})_{ij}=\langle a^{\dagger}_{j}a_{i}\rangle_\mathbf{p}$ for neutrinos ($(\bar{\rho}_{\mathbf{p}})_{ij}=\langle \bar{a}^{\dagger}_{i}\bar{a}_{j}\rangle_\mathbf{p}$ for antineutrinos), for each momentum $\mathbf{p}$ and flavors $i,j$. The diagonal elements are the distribution functions $f_{\nu_{i}\left(\bar{\nu}_{i}\right)}\left(\mathbf{p}\right)$ such that their integration over the momentum space gives the neutrino number density $n_{\nu_{i}}$ of a determined flavor $i$. The off-diagonal elements provide information about the \emph{overlapping} between the two neutrino flavors.

Taking into account the current-current nature of the weak interaction in the standard model, the Hamiltonian for each equation is  \citep{Dolgov:1980cq,Sigl:1992fn,Hannestad:2006nj}
\begin{widetext}
\begin{subequations}
\begin{gather}
H_{\mathbf{p}}=\Omega_{\mathbf{p}}+\sqrt{2}G_{F}\!\!\int\!\!\left( l_{\mathbf{q}}-\bar{l}_{\mathbf{q}}\right)\left( 1-\mathbf{v}_{\mathbf{q}}\cdot\mathbf{v}_{\mathbf{p}} \right)\frac{d^3\mathbf{q}}{\left(2\pi\right)^3}+\sqrt{2}G_{F}\!\!\int\!\!\left( \rho_{\mathbf{q}}-\bar{\rho}_{\mathbf{q}}\right)\left( 1-\mathbf{v}_{\mathbf{q}}\cdot\mathbf{v}_{\mathbf{p}} \right)\frac{d^3\mathbf{q}}{\left(2\pi\right)^{3}}\\
\bar{H}_{\mathbf{p}}=-\Omega_{\mathbf{p}}+\sqrt{2}G_{F}\!\!\int\!\!\left( l_{\mathbf{q}}-\bar{l}_{\mathbf{q}}\right)\left( 1-\mathbf{v}_{\mathbf{q}}\cdot\mathbf{v}_{\mathbf{p}} \right)\frac{d^3\mathbf{q}}{\left(2\pi\right)^3}+\sqrt{2}G_{F}\!\!\int\!\!\left( \rho_{\mathbf{q}}-\bar{\rho}_{\mathbf{q}}\right)\left( 1-\mathbf{v}_{\mathbf{q}}\cdot\mathbf{v}_{\mathbf{p}} \right)\frac{d^3\mathbf{q}}{\left(2\pi\right)^{3}}
\end{gather}\label{eq:FullHam}\end{subequations}\end{widetext}
where $\Omega_{\mathbf{p}}$ is the matrix of vacuum oscillation frequencies, $l_{\mathbf{p}}$ and $\bar{l}_{\mathbf{p}}$ are matrices of occupation numbers for charged leptons built in a similar way to the neutrino matrices, and $\mathbf{v}_{\mathbf{p}}=\mathbf{p}/ p$ is the velocity of a particle with momentum $\mathbf{p}$ (either neutrino or charged lepton).

As in Sec.~\ref{sec:2} we will only consider two neutrino flavors: $e$ and $x=\mu+\tau$. Three-flavor oscillations can be approximated to two-flavor oscillations as a result of the strong hierarchy of the squared mass differences $\vert \Delta m^{2}_{13} \vert \approx \vert \Delta m^{2}_{23} \vert \gg \vert \Delta m^{2}_{12} \vert$ (see Table \ref{tab:tab3}). In this case, only the smallest mixing angle $\theta_{13}$ is considered. We will drop the suffix for the rest of the discussion. Consequently, the relevant oscillations are $\nu_e \rightleftharpoons \nu_x$ and $\bar\nu_e \rightleftharpoons \bar\nu_x$, and each term in the Hamiltonian governing oscillations becomes a 2~$\times$~2 Hermitian matrix. 

\begin{table}
\centering
\begin{tabular}{| l |}
\hline
$ \Delta m^2_{21}  = 7.37\,(6.93-7.97)\times 10^{-5}\T\B$ eV$^2 $ \\
\hline
$|\Delta m^2| = 2.50\,(2.37-2.63) \times 10^{-3}\T\B$ eV$^2$ Normal Hierarchy \\
\hline
$|\Delta m^2| = 2.46 \,(2.33-2.60) \times 10^{-3}\T\B$ eV$^2$ Inverted Hierarchy \\ 
\hline
$\sin^2\theta_{12} = 0.297\,(0.250-0.354)\T\B$ \\
\hline
$\sin^2\theta_{23} (\Delta m^2 > 0) = 0.437\,(0.379-0.616)\T\B$ \\
\hline
$\sin^2\theta_{23} (\Delta m^2 < 0) = 0.569\,(0.383-0.637)\T\B$ \\
\hline
$\sin^2\theta_{13} (\Delta m^2 > 0) = 0.0214\,(0.0185-0.0246)\T\B$ \\
\hline
$\sin^2 \theta_{13} (\Delta m^2 < 0) = 0.0218\,(0.0186-0.0248)\T\B$ \\
\hline
\end{tabular}
\caption{Mixing and squared mass differences as they appear in \cite{Olive:2016xmw}. Error values in parenthesis are shown in 3$\sigma$ interval. The squared mass difference is defined as $\Delta m^2 = m^{2}_{3}-\left(m^{2}_{2}+m^{2}_{1}\right)/2$.}
\label{tab:tab3}
\end{table}

Let us first present the relevant equations for neutrinos. Due to the similarity between $H_{\mathbf{p}}$ and $\bar{H}_{\mathbf{p}}$, the corresponding equations for antineutrinos can be obtained in an analogous manner. In the two-flavor approximation, $\rho$ in Eq.~(\ref{eq:Liouville}) can be written in terms of Pauli matrices and the polarization vector $\mathsf{P}_\mathbf{p}$ as:
\begin{equation}
\small
\rho_{\mathbf{p}}=\left(
 \begin{array}{cc}
  \rho_{ee} & \rho_{ex}\\
  \rho_{xe} & \rho_{xx}\\
   \end{array}\right)_{\mathbf{p}}
   =
 \frac{1}{2}\left(f_{\mathbf{p}}\mathbb{1} +\mathsf{P}_\mathbf{p} \cdot \vec \sigma\right),
	\label{eq:expansion of rho}
\end{equation}
where $f_{\mathbf{p}}={\rm Tr}[\rho_{\mathbf{p}}]=f_{\nu_e}(\mathbf{p})+f_{\nu_x}(\mathbf{p})$ is the sum of the distribution functions for $\nu_e$ and $\nu_x$. Note that the $z$ component of the polarization vector obeys
\begin{equation}
\mathsf{P}^{z}_{\mathbf{p}} =f_{\nu_e}(\mathbf{p})-f_{\nu_x}(\mathbf{p}).
\label{eq:pzeta}
\end{equation}

Hence, this component tracks the fractional flavor composition of the system and appropriately normalizing $\rho_{\mathbf{p}}$ allows to define a survival and mixing probability

\begin{subequations}
\begin{gather}
P_{\nu_{e} \leftrightarrow \nu_{e}} = \frac{1}{2}\left( 1 + \mathsf{P}^{z}_{\mathbf{p}} \right),\\
P_{\nu_{e} \leftrightarrow \nu_{x}} = \frac{1}{2}\left( 1 - \mathsf{P}^{z}_{\mathbf{p}} \right).
\end{gather}\label{eq:survprobability}
\end{subequations}

On the other hand, the Hamiltonian can be written as a sum of three interaction terms:
\begin{equation}
\mathsf{H} = \mathsf{H}_{\mbox{\footnotesize{vacuum}}} + \mathsf{H}_{\mbox{\footnotesize{matter}}} + \mathsf{H}_{\nu\nu}.
\label{neutrinohamiltonian}
\end{equation}
where $\mathsf{H}$ is the two-flavor Hamiltonian. The first term is the Hamiltonian in vacuum~\citep{Qian:1994wh}:
\begin{equation}
\mathsf{H}_{\mbox{\footnotesize{vacuum}}} =\frac{\omega_\mathbf{p}}{2}
\left(
 \begin{array}{cc}
  -\cos 2\theta & \sin 2\theta\\
  \sin 2\theta & \cos 2\theta \\
   \end{array}\right)
   =\frac{\omega_\mathbf{p}}{2} \vec{B}\cdot \vec{\sigma}
	\label{Hvacuum}
\end{equation}
where $\omega_\mathbf{p} = \Delta m^2/2p$, $\vec{B}=(\sin2\theta,0,-\cos 2 \theta)$ and $\theta$ is the smallest neutrino mixing angle in vacuum. 

The other two terms in Eqs.~(\ref{eq:FullHam}) are special since they make the evolution equations non-linear. Even though they are very similar, we are considering that the electrons during the accretion form an isotropic gas; hence, the vector $\mathbf{v}_{\mathbf{q}}$ in the first integral is distributed uniformly on the unit sphere and the factor $\mathbf{v}_\mathbf{q}\cdot\mathbf{v}_\mathbf{p}$ averages to zero. After integrating the matter Hamiltonian is given by:
\begin{equation}
\mathsf{H}_{\mbox{\footnotesize{matter}}} =
\frac{\lambda}{2}\left(
 \begin{array}{cc}
  1 & 0\\
  0 & -1 \\
   \end{array}\right)
   =\frac{\lambda}{2} \vec{L} \cdot \vec{\sigma}
	\label{Hmatter}
\end{equation}
where $\lambda = \sqrt{2}G_{F}\left(n_{e^-} - n_{e^+}\right)$ is the charged current matter potential and $\vec{L}=(0,0,1)$.

Such simplification cannot be made with the final term. Since neutrinos are responsible for the energy loss of the infalling material during accretion, they must be escaping the accretion zone and the net neutrino and antineutrino flux is non-zero.In this case the factor $\mathbf{v}_\mathbf{q}\cdot\mathbf{v}_\mathbf{p}$ cannot be averaged to zero. At any rate, we can still use Eq.~(\ref{eq:expansion of rho}) and obtain \citep{1992PhLB..287..128P,2016arXiv160704671Z,2016PhRvD..93d5021M}:
\begin{equation}
\mathsf{H}_{\nu\nu} = \sqrt{2}G_{F}\left[ \int\!\! \left(1- \mathbf{v}_{\mathbf{q}}\cdot\mathbf{v}_{\mathbf{p}}\right) \left(\mathsf{P}_\mathbf{q}-\bar{\mathsf{P}}_\mathbf{q}\right)\frac{d^3\mathbf{q}}{\left(2\pi\right)^3}\right]\cdot \vec{\sigma}
\label{Hnunu}
\end{equation}

Introducing every Hamiltonian term in Eqs.~(\ref{eq:Liouville}), and using the commutation relations of the Pauli matrices, we find the EoM for neutrinos and antineutrinos for each momentum mode $\mathbf{p}$:

\begin{subequations}
\begin{gather}
\dot{\mathsf{P}}_\mathbf{p} = \left[ \omega_\mathbf{p} \vec{B} + \!\lambda \vec{L} + \!\! \sqrt{2}G_{F}\!\!\!  \int\!\! \left(1- \mathbf{v}_{\mathbf{q}}\!\cdot\mathbf{v}_{\mathbf{p}}\right) \left(\mathsf{P}_\mathbf{q}-\bar{\mathsf{P}}_\mathbf{q}\right)\frac{d^3\mathbf{q}}{\left(2\pi\right)^3} \right] \times \mathsf{P}_\mathbf{p}\\
\dot{\bar{\mathsf{P}}}_\mathbf{p} = \left[ -\omega_\mathbf{p} \vec{B} + \!\lambda \vec{L} + \!\! \sqrt{2}G_{F}\!\!\!  \int\!\! \left(1- \mathbf{v}_{\mathbf{q}}\!\cdot\mathbf{v}_{\mathbf{p}}\right) \left(\mathsf{P}_\mathbf{q}-\bar{\mathsf{P}}_\mathbf{q}\right)\frac{d^3\mathbf{q}}{\left(2\pi\right)^3}\right]\times \bar{\mathsf{P}}_\mathbf{p}.\end{gather}\label{eq:Hnu}\end{subequations}

Solving the above equations would yield the polarization vectors as a function of time. However, in our specific physical system, both the matter potential $\lambda$ and the neutrino potential vary with the radial distance from the NS surface as well as the instant $t$ of the physical process which can be characterized by the accretion rate $\dot{M}$. As we will see later, the time dependence can be ignored. This means that Eqs.~(\ref{eq:Hnu}) must be written in a way that makes explicit the spatial dependence, i.e. in terms of the vector $\mathbf{r}$. For an isotropic and homogeneous neutrino gas or a collimated ray of neutrinos the expression $dt=dr$ would be good enough, but for radiating extended sources the situation is more complicated. In Eqs.~(\ref{eq:Liouville}) we must replace the matrices of occupation numbers by the space dependent Wigner functions $\rho_{\mathbf{p,r}}$ (and $\bar{\rho}_{\mathbf{p,r}}$) and the total time derivative by the Liouville operator \citep{Cardall:2007zw,Strack:2005ux}
\begin{equation}
\dot{\rho}_{\mathbf{p,r}}=\frac{\partial \rho_{\mathbf{p,r}}}{\partial t} + \mathbf{v}_\mathbf{p} \cdot \nabla_{\mathbf{r}}\,\rho_{\mathbf{p,r}} + \dot{\mathbf{p}}\cdot\nabla_{\mathbf{p}}\,\rho_{\mathbf{p,r}}
	\label{Loperator}
\end{equation}

We will ignore the third term of the Liouville operator since we won't consider the gravitational deflection of neutrinos. For peak accretion rates $\dot{M} \approx 10^{-8}$--$10^{-2}~M_{\odot}/{\rm s}$ the characteristic accretion time is $\Delta t_{acc}= M/\dot{M}\approx M_\odot/\dot{M}\approx 10^8$--$10^2$~s. The distances traveled by a neutrino in these times are $r \approx 3\times 10^{12}$--$3\times 10^{18}$~cm. These distances are much larger than the typical binary separation $a$. As a consequence, we can consider the neutrino evolution to be a stationary process. This fact allows us to neglect the first term in Eq.~(\ref{Loperator}). Putting together these results, the EoM become:
\begin{subequations}
\begin{gather}
i\mathbf{v}_\mathbf{p} \cdot \nabla_{\mathbf{r}}\,\rho_{\mathbf{p,r}} =
[H_{\mathbf{p,r}},\rho_{\mathbf{p,r}}]\\
i\mathbf{v}_\mathbf{p} \cdot \nabla_{\mathbf{r}}\,\bar{\rho}_{\mathbf{p,r}} =
[\bar{H}_{\mathbf{p,r}},\bar{\rho}_{\mathbf{p,r}}],	\end{gather}\label{eq:Lspace1}\end{subequations}
where ${H}_{\mathbf{p,r}}$ and $\bar{H}_{\mathbf{p,r}}$ are the same as (\ref{eq:FullHam}) but the matrices of densities (as well as the polarization vectors) depend on the position $\mathbf{r}$. Note, however, that the electrons in the accretion zone still form an isotropic gas and Eq.~(\ref{Hmatter}) is still valid and the matter Hamiltonian depends on $\mathbf{r}$ through $n_{e^{-}}(\mathbf{r})-n_{e^{+}}(\mathbf{r})$. The first two terms in the Hamiltonian remain virtually unchanged. On the other hand, projecting the EoM onto the radial distance from the NS and using the axial symmetry of the system, the integral in the neutrino-neutrino interaction term can be written as
\begin{equation}
\frac{\sqrt{2}G_{F}}{\left(2\pi\right)^{2}}\int\!\! \left(1- v_{\vartheta^{\prime}_{r}}v_{\vartheta_{r}}\right)\left( \rho_{q,\vartheta^{\prime}\!,r}\! - \bar{\rho}_{q,\vartheta^{\prime}\!,r} \right)q^{2}\! dq \vert d\cos\vartheta^{\prime}_{r} \vert.
\label{eq:n-ninterterm}
\end{equation}

Since the farther from the NS the interacting neutrinos approach a perfect collinearity, the projected velocities $v_{\vartheta_{r}}$ become decreasing functions of the position. In this particular geometry the diagonal elements of the matrix of densities are written as a product of independent distributions over each variable $p,\vartheta,\phi$, where the $\phi$ dependence has been integrated out. The one over $p$ is the normalized Fermi-Dirac distribution and the one over $\vartheta$ is assumed uniform due to symmetry. The $r$ dependence is obtained through the geometrical flux dilution. Knowing this, the diagonal elements of matrices of densities at the NS surface are
\begin{subequations}
\begin{gather}
\left(\rho_{\mathbf{p,R_{\rm NS}}} \right)_{ee} = \left(\bar{\rho}_{\mathbf{p,R_{\rm NS}}}\right)_{ee} =  f_{\nu_{e}}\!\left(\mathbf{p}\right)\\
\left(\rho_{\mathbf{p,R_{\rm NS}}} \right)_{xx} = \left(\bar{\rho}_{\mathbf{p,R_{\rm NS}}}\right)_{xx} = f_{\nu_{x}}\!\left(\mathbf{p}\right) \end{gather}\label{eq:distributionfactoring1}\end{subequations}
where the functions $f_{\nu_{i}}$ are given by Eqs.~(\ref{eq:neutrinofullinitialspectrum}).

\subsection{Single-angle approximations}\label{sub:single}
The integro-differential Eqs.~(\ref{eq:Hnu}) and (\ref{eq:Lspace1}) are usually numerically solved for the momentum $\mathbf{p}$ and the scalar $\mathbf{v}_{\mathbf{q}}\cdot\mathbf{v}_{\mathbf{p}}$. Such simulation are quite time-consuming and the result is frequently too complicated to allow for a clear interpretation of the underlying physics. For this reason, the analytic approximation called the \emph{single-angle limit} is made. Such approximation consists in \emph{imposing} a self-maintained coherence in the neutrino system, i.e. it is assumed that the flavor evolution of all neutrinos emitted from an extended source is the same as the flavor evolution of the neutrinos emitted from the source along a particular path. Under this premise, the propagation angle between the test neutrino and the background neutrinos is fixed. In expression (\ref{eq:n-ninterterm}) this is equivalent to dropping the $\vartheta^{\prime}$ dependence of $\rho$ and replacing the projected velocity $v_{\vartheta_r}$ either by an appropriate average at each $r$ \citep[as in][]{Dasgupta:2007ws} or by a representative angle (usually $0$ or $\pi/4$). We will follow the former approach and apply the \emph{bulb model} described in \citet{Duan:2006an}. Within this model it is shown that the projected velocity at a distance $r$ from the neutrino emission zone is
\begin{equation}
v_{r} = \sqrt{1-\left(\frac{R_{\rm NS}}{r}\right)^{2}\left( 1-v^{2}_{R_{\rm NS}} \right)}.
\label{eq:projvelocity}
\end{equation}
where $v_{R_{\rm NS}}$ is the projected velocity at the NS surface. By redefining the matrices of density with a change of variable $u=1-v^{2}_{R_{\rm NS}}$ in the integral (\ref{eq:n-ninterterm})
\begin{equation}
\rho_{p,u,r}\frac{p^{2}}{2\left(2 \pi \right)^{2}} \rightarrow \rho_{p,u,r},
\label{eq:matricesforflux}
\end{equation}
and using Eq.~(\ref{eq:expansion of rho}), we can write the full equations of motion
\begin{subequations}
\begin{gather}
\frac{\partial}{\partial r} \mathsf{P}_{p,r}\! = \! \left[\omega_{p,r}\vec{B} + \!\lambda_{r}\vec{L} + \mu_{r}\!\!\! \int^{\infty}_{0}\!\!\! \left(\mathsf{P}_{q,r}\! - \bar{\mathsf{P}}_{q,r}\right) dq \right] \! \times \! \mathsf{P}_{p,r} \\
\frac{\partial}{\partial r} \bar{\mathsf{P}}_{p,r}\! = \! \left[-\omega_{p,r}\vec{B} +\! \lambda_{r}\vec{L} + \mu_{r}\!\!\! \int^{\infty}_{0}\!\!\! \left(\mathsf{P}_{q,r}\! - \bar{\mathsf{P}}_{q,r}\right) dq \right] \! \times \! \bar{\mathsf{P}}_{p,r}\end{gather}\label{eq:fulleqinspace1}\end{subequations}
where we have replaced $v_{r}$ by it's average value
\begin{equation}
\langle v_{r} \rangle = \frac{1}{2}\left[ 1+\sqrt{1 - \left(\frac{R_{\rm NS}}{r}\right)^{2}} \right].
\label{eq:averageradialvelocity}
\end{equation}

All the interaction potentials now depend on $r$ and each effective potential strength is parametrized as follows \citep{Dasgupta:2007ws}
\begin{equation}
\omega_{p,r} =\! \frac{\Delta m^{2}}{2p\langle v_{r} \rangle},
\label{eq:vacuumpotential}
\end{equation}
\begin{equation}
\lambda_{r} \!=\! \sqrt{2}G_{F}\left(n_{e^{-}}(r)\!-n_{e^{+}}(r)\right)\frac{1}{\langle v_{r} \rangle},
\label{eq:matterpotential}
\end{equation}
\begin{equation}
\mu_{r} \!=\!\frac{ \sqrt{2}G_{F}}{2}\left(\sum_{i\in\{e,x\}}\!n^{C}_{\nu_{i}\bar{\nu}_{i}}\right)\left( \frac{R_{\rm NS}}{r} \right)^{2}\left( \frac{1 - \langle v_{r} \rangle^{2}}{\langle v_{r} \rangle} \right).
\label{eq:neutrinopotential}
\end{equation}

It is worth mentioning that all the effective potential strengths are affected by the geometry of the extended source through the projected velocity on the right side of Eqs.~(\ref{eq:Lspace1}).
For the neutrino-neutrino interaction potential, we have chosen the total neutrino number density as parametrization. This factor comes from the freedom to re-normalize the polarization vectors in the EoM. A different choice has been made in \citet{EstebanPretel:2007ec}. Of the other two $r$ dependent factors, one comes from the geometrical flux dilution and the other accounts for collinearity in the single-angle approximation. Over all $\mu_{r}$ decays as $1/r^{4}$.

\begin{figure*}
\centering
\includegraphics[width=0.49\hsize,clip]{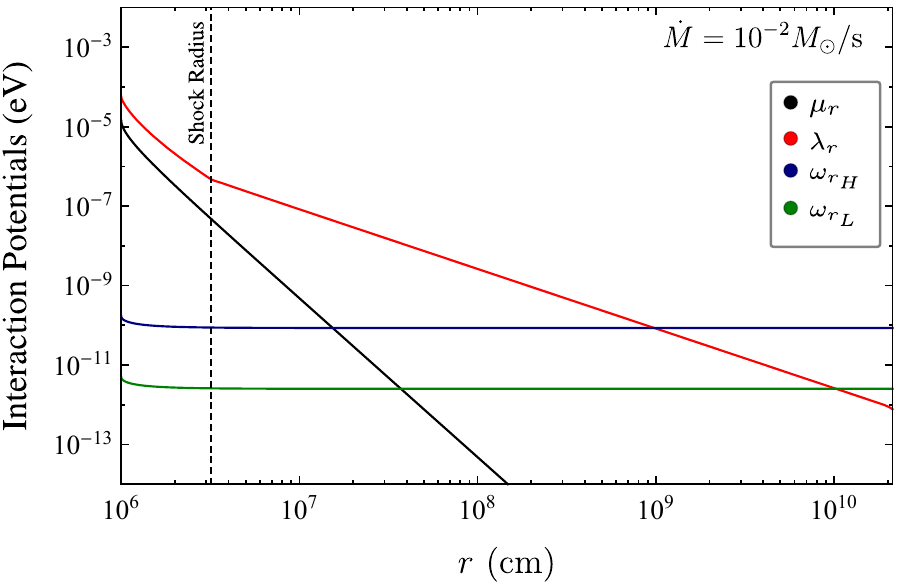}\includegraphics[width=0.49\hsize,clip]{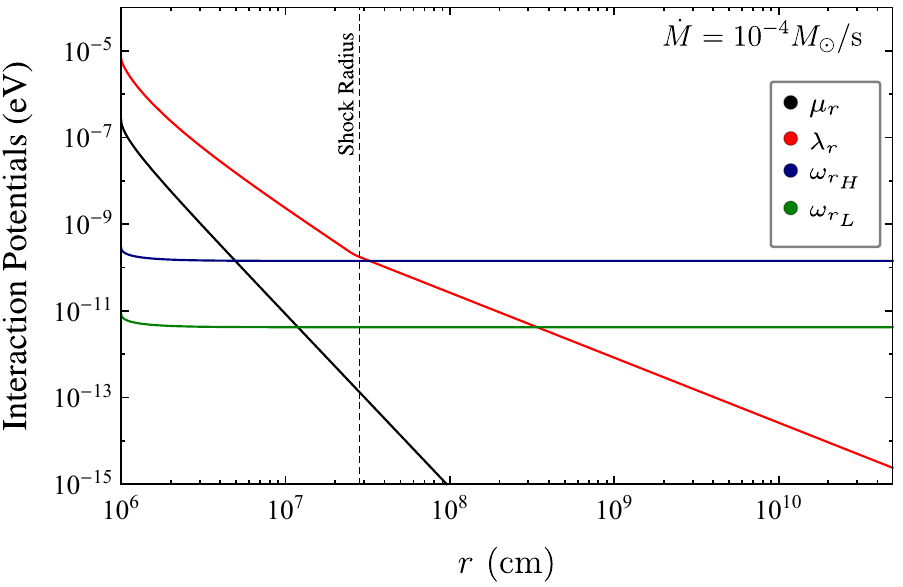}\\
\includegraphics[width=0.49\hsize,clip]{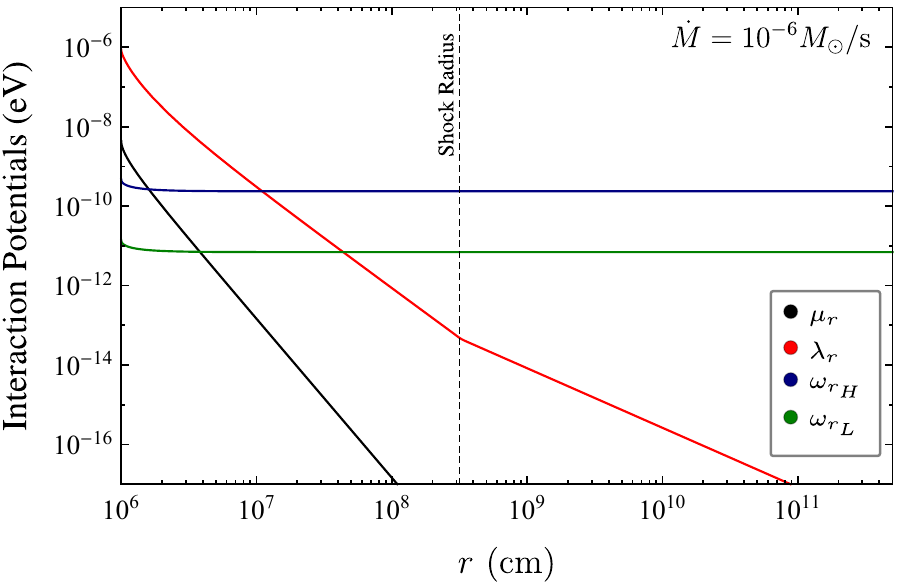}\includegraphics[width=0.49\hsize,clip]{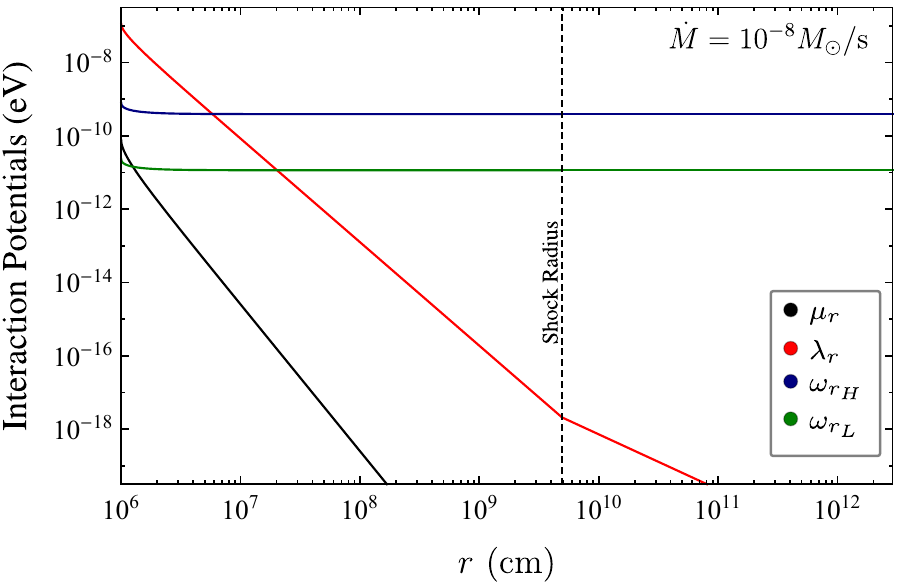}\caption{Interaction potentials as functions of the radial distance from the NS center for selected accretion rates $\dot{M}$ (see Table \ref{tab:tab1}). Each plot runs from the NS surface to the Bondi-Hoyle surface. $\mu_{r}$ stands for the self-interaction neutrino potential, $\lambda_{r}$ is the matter potential and $\omega_H$ and $\omega_L$ are the higher and lower resonances corresponding to the atmospheric and solar neutrino scales, respectively, defined in Eq.~(\ref{resonances}). Outside the Bondi-Hoyle region the neutrino and electron densities depend on the direction of their path relative to the SN and the particular ejecta density profile.} 
\label{fig:potentials}
\end{figure*}

\begin{figure*}
\centering
\textbf{\large Inverted Hierarchy}\par\medskip
\includegraphics[width=0.44\hsize,clip]{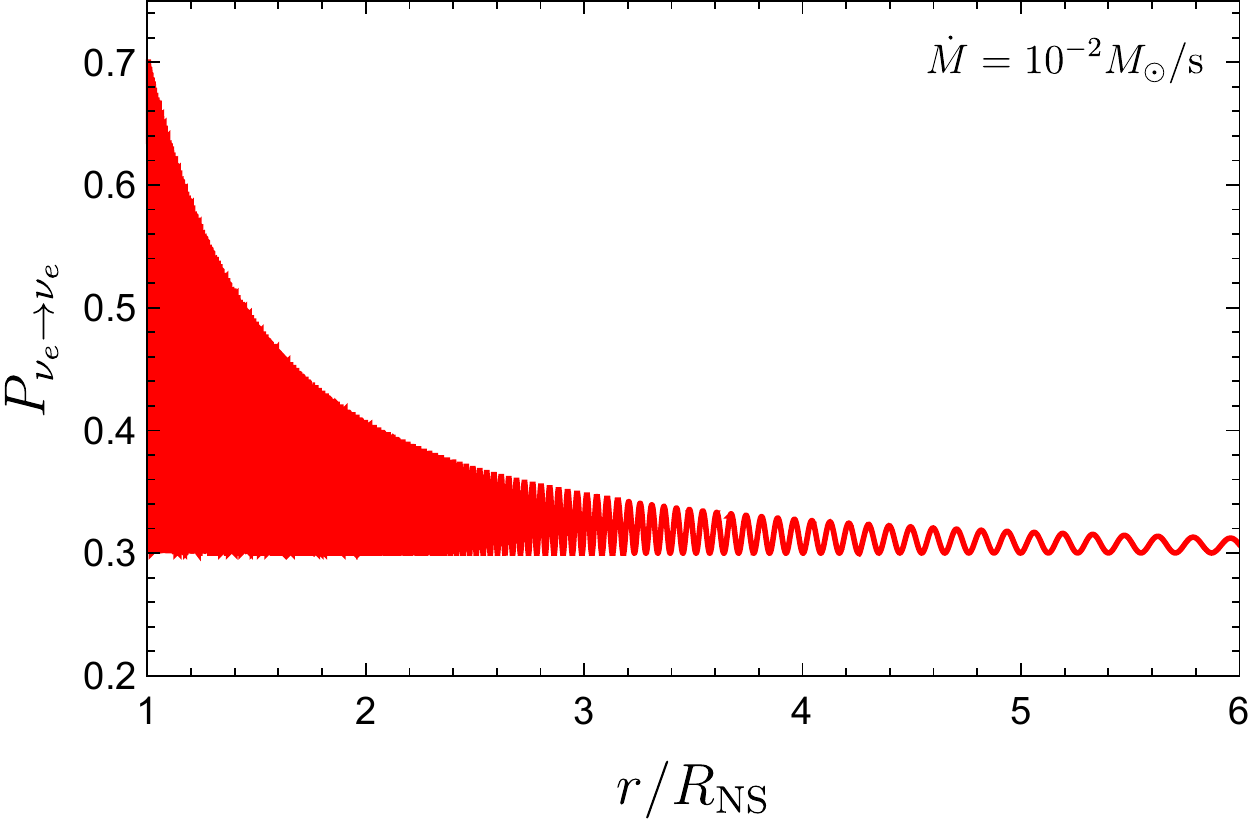}\includegraphics[width=0.44\hsize,clip]{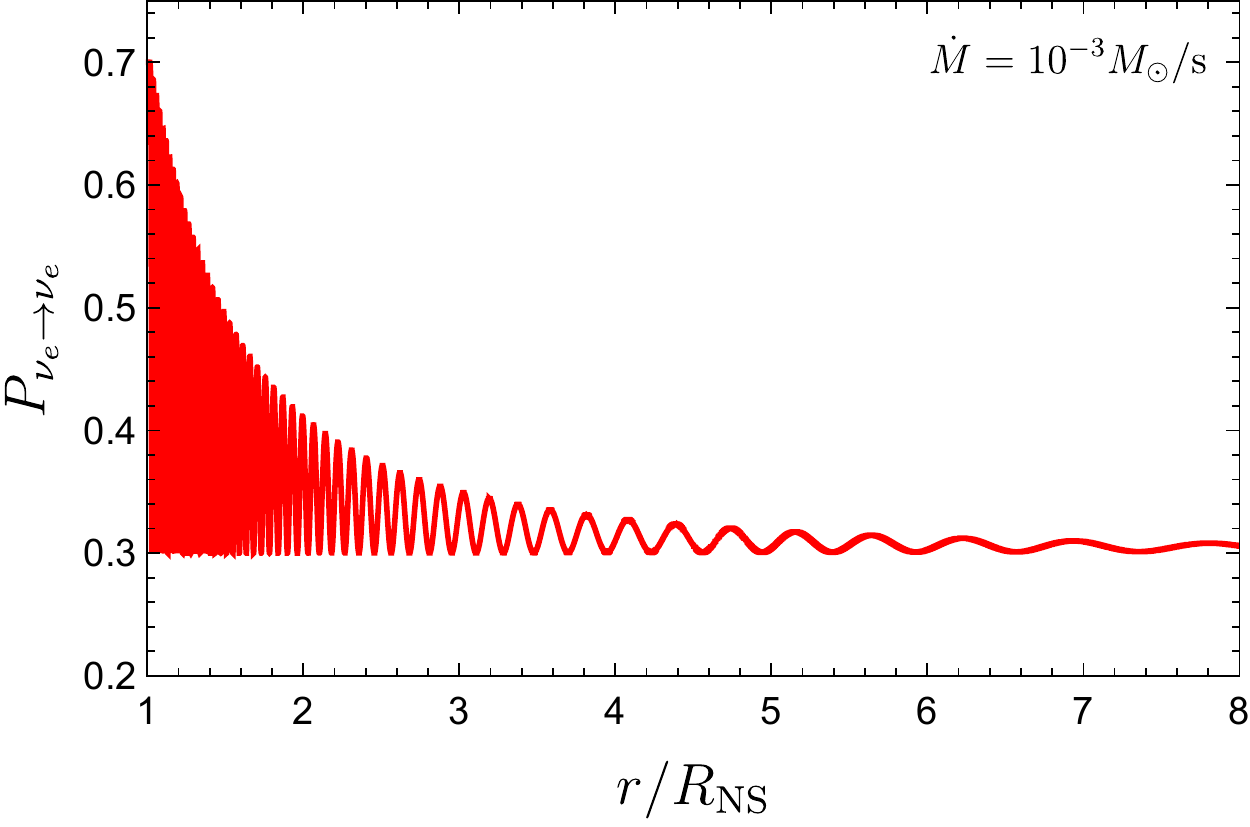}\\
\includegraphics[width=0.44\hsize,clip]{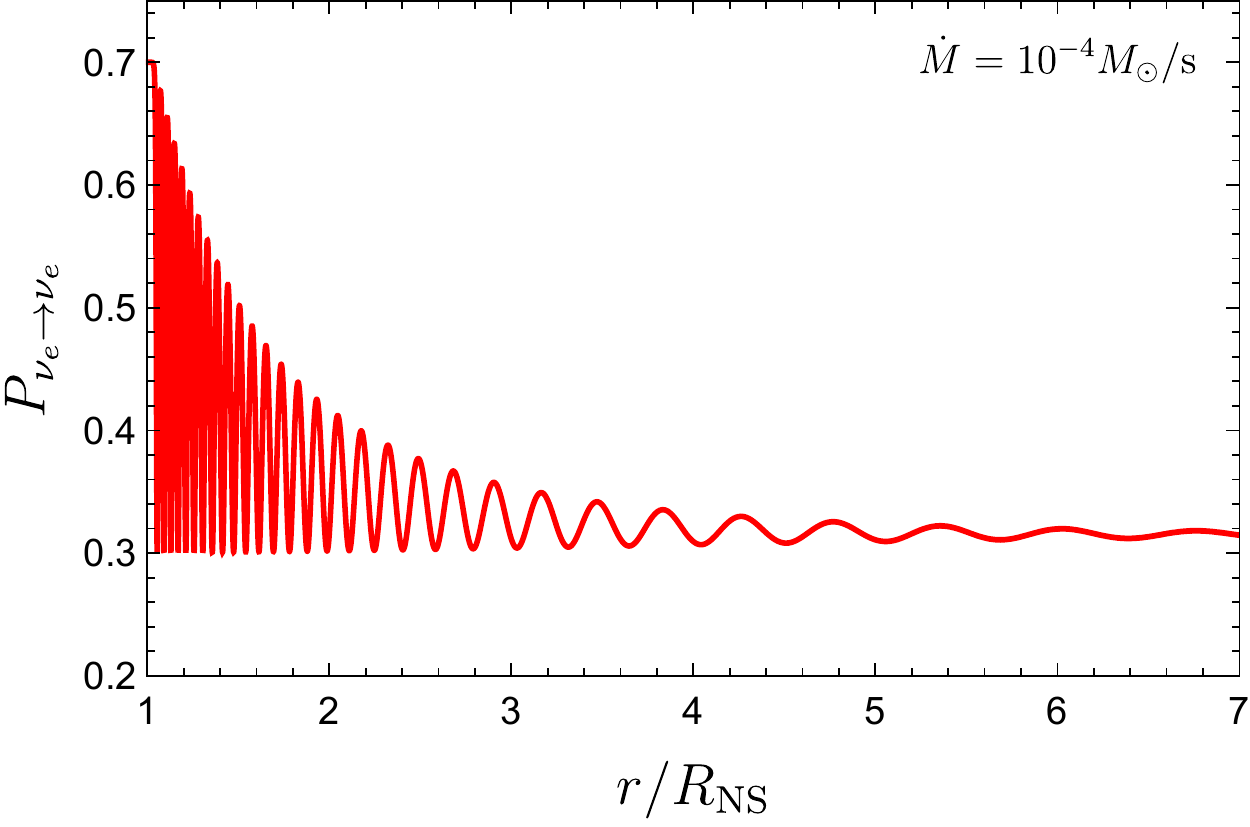}\includegraphics[width=0.44\hsize,clip]{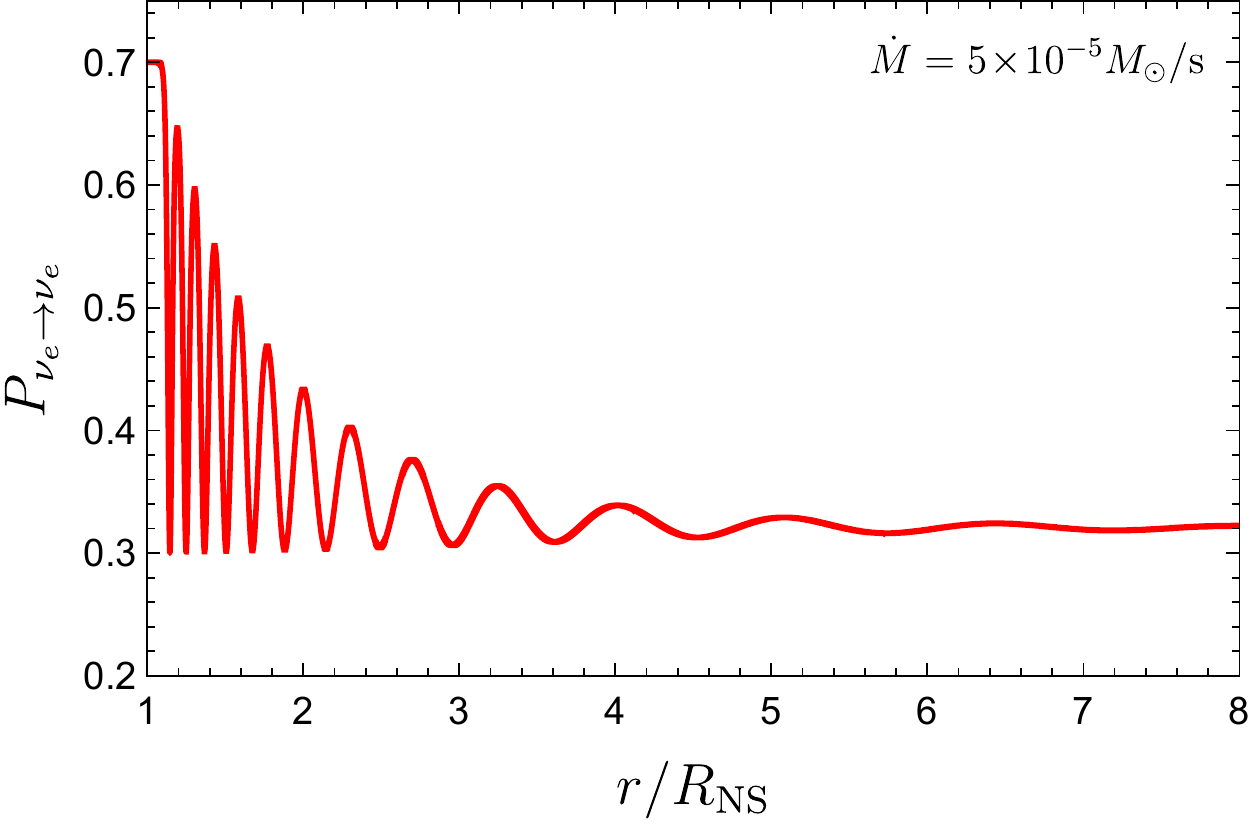}
\caption{Neutrino flavor evolution for inverted hierarchy. Electron neutrino survival probability is shown as a function of the radial distance from the NS surface. The curves for the electron antineutrino match the ones for electron neutrinos.} 
\label{fig:singleangle}
\end{figure*}
\begin{figure*}
\centering
\textbf{\large Normal Hierarchy}\par\medskip
\includegraphics[width=0.44\hsize,clip]{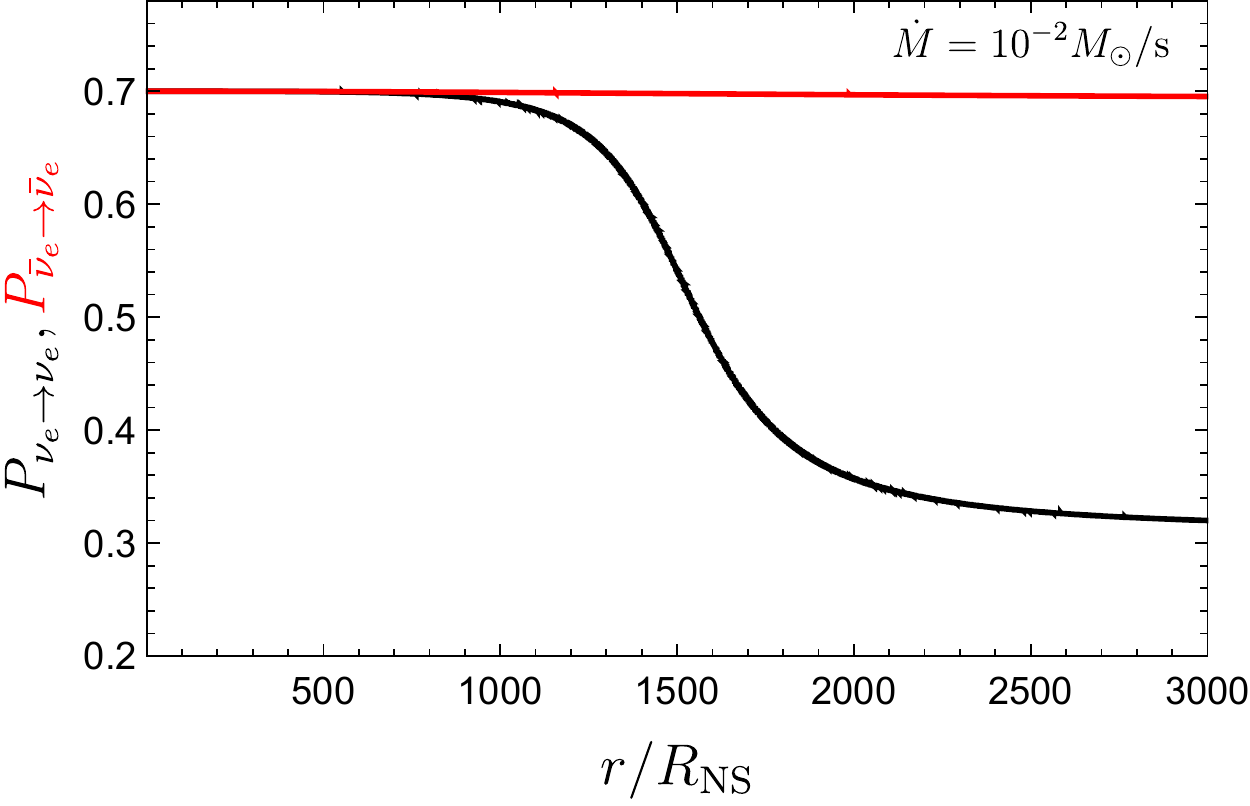}\includegraphics[width=0.44\hsize,clip]{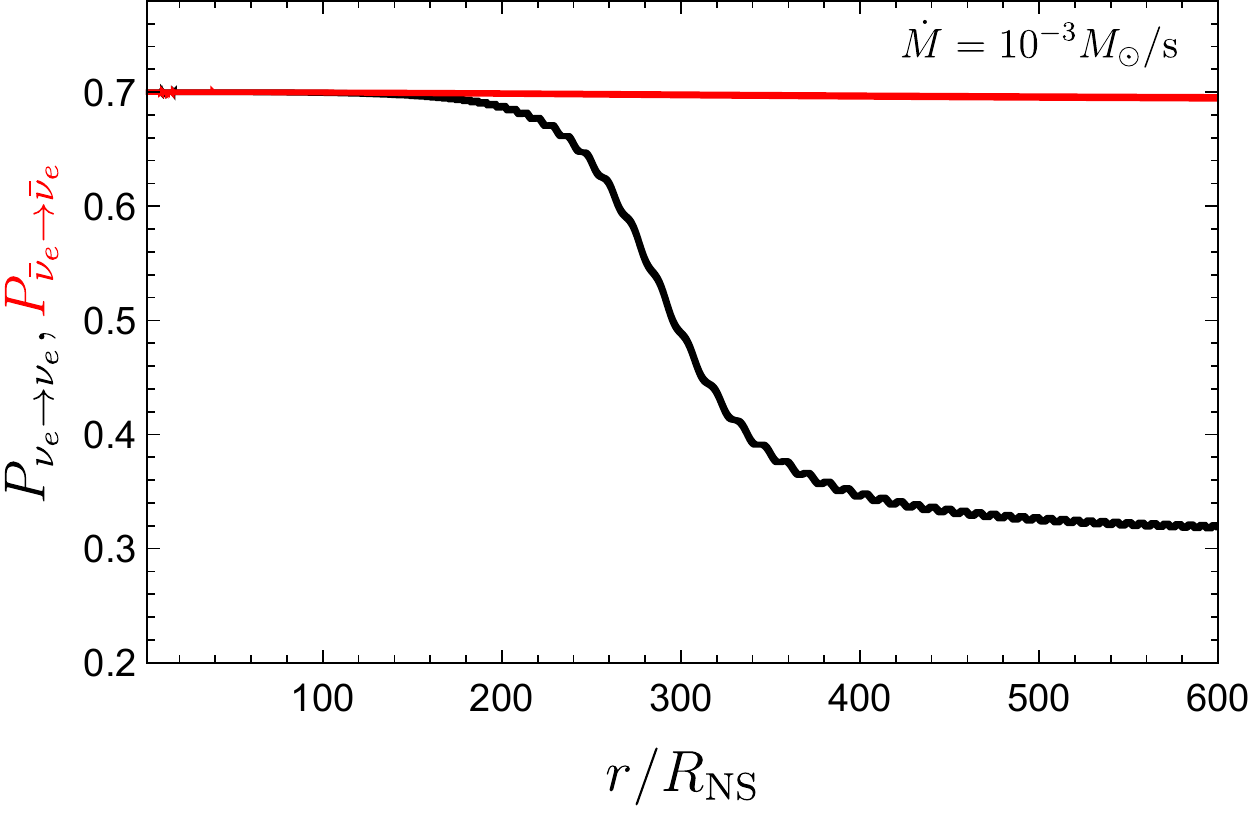}\\
\includegraphics[width=0.44\hsize,clip]{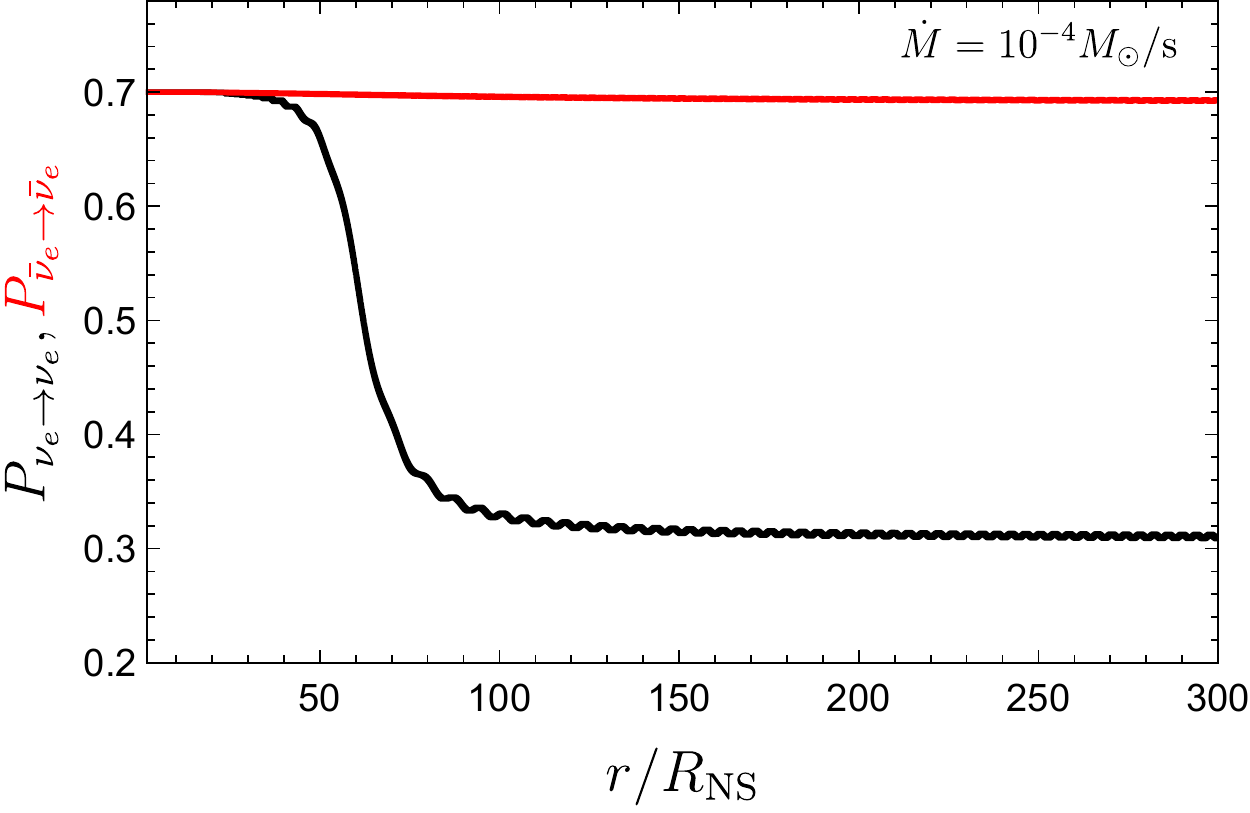}\includegraphics[width=0.44\hsize,clip]{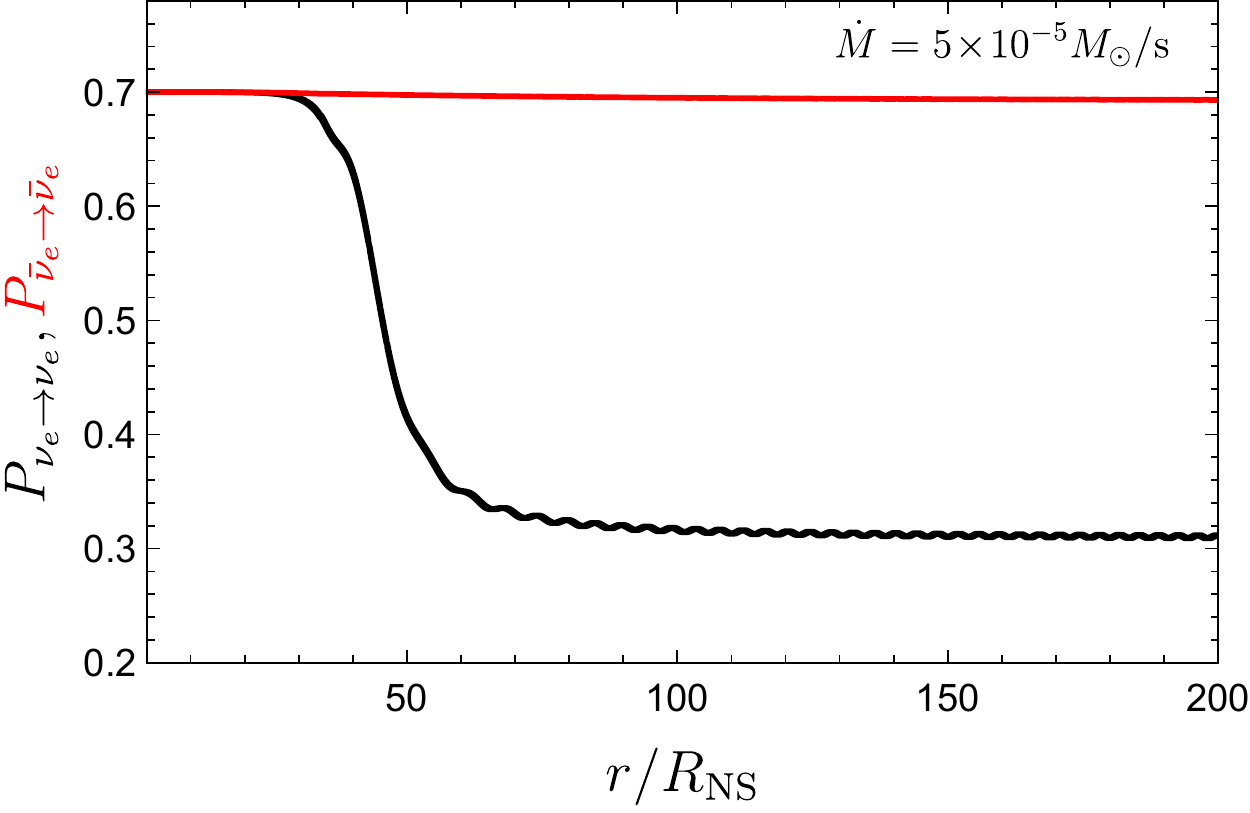}
\caption{Electron neutrino and antineutrino flavor evolution for normal hierarchy.The survival probability is shown as a function of the radial distance from the NS surface.} 
\label{fig:singleanglesolutions}
\end{figure*}

In Fig.~\ref{fig:potentials} the behavior of the effective potentials within the single-angle formalism is shown for $\dot{M}= 10^{-2} M_\odot$~s$^{-1}$, $10^{-4} M_\odot$~s$^{-1}$, $10^{-6} M_\odot$~s$^{-1}$ and $10^{-8} M_\odot$~s$^{-1}$. In all cases, the neutrino energy is the corresponding average reported in Table \ref{tab:tab1}. Since the oscillatory dynamics of the neutrino flavors are determined by the value of the potentials, and the value of the potentials depends on the data in Table \ref{tab:tab1}, it is important to establish how sensible is this information to the model we have adopted. In particular, to the pre-SN envelope density profile index $m$. The reported accretion rates can be seen as different states in the evolution of a binary system or as peak accretion rates of different binary systems. For a given accretion rate, the temperature and density conditions on the neutron star surface are fixed. This, in turn, fixes the potentials involved in the equations of flavor evolution and the initial neutrino and antineutrino flavor content. To see the consequences of changing the index $m$ we can estimate the peak accretion rates for new values using Eqs.~(\ref{eq:mpeakandtpeak}). Since we are only interested in type Ic supernovae, we shall restrict these values to the ones reported in Table 1 of \citet{2016ApJ...833..107B} (that is $m=2.771$, $2.946$ and $2.801$), and in each case, we consider the smallest binary separation such that there is no Roche-Lobe overflow. For these parameters, we find peak accretion rates $\dot{M}_{\rm peak} \sim 10^{-2}$--$10^{-4} M_\odot$~s$^{-1}$ with peak times at $t_{\rm peak} \approx 7$--$35$ min. Because these accretion rates are still within the range in Table \ref{tab:tab1}, the results contained in Sec.~\ref{sec:solutions} apply also to these cases with different value of the $m$-index.

The profiles for the electron and positron number densities were adopted from the simulations presented in \cite{2016ApJ...833..107B}. Due to the dynamics of the infalling matter, close to the NS, the behavior of $n_{e^{-}}(r) -n_{e^{+}}(r)$ is similar to $\mu_{r}$. At the shock radius, the electron density's derivative presents a discontinuity and its behavior changes allowing for three distinct regions inside the Bondi-Hoyle radius. The matter potential is always higher than the neutrino potential yet, in most cases, both are higher than the vacuum potential, so we expect neutrino collective effects (neutrino oscillations) and MSW resonances to play a role in the neutrino flavor evolution inside the Bondi-Hoyle radius. Outside the capture region, as long as the neutrinos are not directed towards the SN, they will be subjected to vacuum oscillations.
%

\section{Single-angle solutions and multi-angle effects}\label{sec:solutions}

%
The full dynamics of neutrino oscillations is a rather complex interplay between the three potentials discussed in Sec.~\ref{sec:oscillations}, yet the neutrino-antineutrino symmetry allows us to generalize our single-angle calculations for certain accretion rates using some numerical and algebraic results obtained in~\cite{Hannestad:2006nj,Fogli:2007bk,EstebanPretel:2007ec} and references therein. Specifically, we know that if $\mu_{r} \gg \omega_{r}$, as long as the MSW condition $\lambda_{r} \simeq \omega_{r}$ is not met, collective effects should dominate the neutrino evolution even if $\lambda_{r} \gg \mu_{r}$. On the other hand, if $\mu_{r} \lesssim \omega_{r}$, the neutrino evolution is driven by the relative values between the matter and vacuum potentials. With this in mind, we identify two different ranges of values for the accretion rate: $\dot{M} \gtrsim 5 \times 10^{-5}M_{\odot}$~s$^{-1}$ and $\dot{M} \lesssim 5 \times 10^{-5}M_{\odot}$~s$^{-1}$.
\subsection{High accretion rates}\label{sec:highar}

For accretion rates $\dot{M} \gtrsim 5 \times 10^{-5}M_{\odot}$~s$^{-1}$ the potentials obey the following hierarchy
\begin{equation}
\lambda_{r} \gtrsim \mu_{r} \gg \omega_{r},
\label{neutrino dominance}
\end{equation}
hence, we expect strong effects of neutrino self-interactions. In order to appreciate the interesting physical processes which happen with the neutrinos along their trajectory in the accretion zone, we begin this analysis with a simplified approach to the EoM for a monochromatic spectrum with the same energy for both neutrinos and antineutrinos. Let us introduce the following definitions
\begin{equation}
\vec{D}=\mathsf{P}_{r}-\bar{\mathsf{P}}_{r}
\label{definition D}
\end{equation}
\begin{equation}
\vec{Q}=\mathsf{P}_{r} + \bar{\mathsf{P}}_{r} - \frac{\omega_{r}}{\mu_{r}}\vec{B}.
\label{definitions Q}
\end{equation}

The role of the matter potential is to logarithmically extend the period of the bipolar oscillations so we can ignore it for now. Also, we will restrict our analysis to a small enough region at $R_{\rm NS}+\Delta r_{\nu}$ so that we can consider $\frac{d}{dr}(\omega_{r}/\mu_{r}) \approx 0$ (adiabatic approximation). Then, By summing and subtracting Eqs.~(\ref{eq:fulleqinspace1}) and using definitions (\ref{definition D}) and (\ref{definitions Q}), we obtain
\begin{equation}
\frac{d}{d r}\vec{Q}= \mu \vec{D}\times \vec{Q}
\label{evolution Q}
\end{equation}
\begin{equation}
\frac{d}{d r} \vec{D}= \omega \vec{B} \times \vec{Q}.
\label{evolution D}
\end{equation}

We are now able to build a very useful analogy. The equations above are analogous to the EoM of a simple mechanical pendulum with a vector position given by $\vec{Q}$, precessing around an angular momentum $\vec{D}$, subjected to a force $\omega\mu\vec{B}$ with a moment of inertia proportional to the inverse of $\mu$. With Eqs.~(\ref{eq:creationdensityflux}) and (\ref{eq:pzeta}) the initial conditions for the polarization vectors are
\begin{eqnarray}
\mathsf{P}(R_{\rm NS}) = \bar{\mathsf{P}}(R_{\rm NS}) = (0,0,0.4)
\label{eq:initialconditions}
\end{eqnarray}

We can easily show that $\vert\vec{Q}(R_{\rm NS})\vert = \vert\mathsf{P}(R_{\rm NS})+\bar{\mathsf{P}}(R_{\rm NS}) \vert + O(\omega / \mu) \approx 0.8$. Calculating $\frac{d}{dr}(\vec{Q}\cdot\vec{Q})$ it can be checked that this value is conserved.

The analogous angular momentum is $\vec{D}(R_{\rm NS})=\mathsf{P}(R_{\rm NS})-\bar{\mathsf{P}}(R_{\rm NS}) = 0$. Thus, the pendulum moves initially in a plane defined by $\vec{B}$ and the $z$-axis, i.e., the plane $xz$. Then, it is possible to define an angle $\varphi$ between $\vec{Q}$ and the $z$-axis such that
\begin{equation}
\vec{Q}=\vert\vec{Q}|\left(\sin\varphi,0,\cos\varphi\right).
\label{definiton phi}
\end{equation}
Note that the only non-zero component of $\vec{D}$ is $y$-component and from (\ref{evolution Q}) and (\ref{evolution D}) we find
\begin{equation}
\frac{d\varphi}{d r} = \mu \vert\vec{D}\vert
\label{evolution phi1}
\end{equation}
and
\begin{equation}
\frac{d\vert\vec{D}\vert}{d r} = - \omega \vert\vec{Q}\vert \cos (2\theta + \varphi).
\label{evolution phi2}
\end{equation}

The above equations can be equivalently written as
\begin{equation}
\frac{d^2 \varphi}{d r^2} = - k^2 \sin (2\theta + \varphi),
\label{motion of phi}
\end{equation}
where we have introduced the inverse characteristic distance $k$ by
\begin{equation}
k^2=\omega \mu\vert\vec{Q}\vert,
\label{deffinition k}
\end{equation}
which is related to the anharmonic oscillations described by the non-linear EoM (\ref{evolution phi1}) and (\ref{evolution phi2}). The logarithmic correction to the oscillation length due to matter effects is \citep{Hannestad:2006nj}
\begin{equation}
\tau_{\dot{M}} =-k^{-1}\ln\left[\left(\frac{\pi}{2}-\theta\right)\frac{k}{\left(k^{2}+\lambda^{2}\right)^{1/2}}\left(1 + \frac{\omega}{\vert\vec{Q}\vert\mu}\right)\right].
\label{deffinition tau}
\end{equation}
The initial conditions (\ref{eq:initialconditions}) imply
\begin{equation}
\varphi\left(R_{\rm NS}\right) = \arcsin \left( \frac{\omega}{\vert\vec{Q}\vert\mu}\sin 2\theta \right).
\label{initial phi}
\end{equation}

To investigate the physical meaning of the above equation, let us assume for a moment that $2\theta$ is a small angle. In this case $\varphi\left(R_{\rm NS}\right)$ is also a small angle. If $k^2>0$, which is true for the normal hierarchy $\Delta m^{2}  > 0$, we expect small oscillations around the initial position since the system begins in a stable position of the potential associated with Eqs.~(\ref{evolution phi1}) and (\ref{evolution phi2}). No strong flavor oscillations are expected. On the contrary, for the inverted hierarchy $\Delta m^{2} < 0$, $k^{2}<0$ and the initial $\varphi(R_{\rm NS})$ indicates that the system begins in an unstable position and we expect very large anharmonic oscillations. $\mathsf{P}^{z}$ (as well as $\bar{\mathsf{P}}^{z}$) oscillates between two different maxima passing through a minimum $-\mathsf{P}^{z}$ ($-\bar{\mathsf{P}}^{z}$) several times. This behavior implies total flavor conversion: all electronic neutrinos (antineutrinos) are converted into non-electronic neutrinos (antineutrinos) and vice-versa. This has been called bipolar oscillations in the literature~\citep{Duan:2010bg}. 

We solved numerically Eqs.~(\ref{eq:fulleqinspace1}) for both normal and inverted hierarchies using a monochromatic spectrum dominated by the average neutrino energy for $\dot{M}=10^{-2},10^{-3},10^{-4}$ and $5 \times 10^{-5} M_{\odot}$~s$^{-1}$, and the respective values reported in Table \ref{tab:tab1} with the initial conditions given by Eqs.~(\ref{eq:creationdensityflux}) and (\ref{eq:distributionfactoring1}). The behavior of the electronic neutrino survival probability inside the accretion zone is shown in Figs.~\ref{fig:singleangle} and \ref{fig:singleanglesolutions} for inverted hierarchy and normal hierarchy, respectively. For the inverted hierarchy, there is no difference between the neutrino and antineutrino survival probabilities. This should be expected since for these values of $r$ the matter and self-interaction potentials are much larger than the vacuum potential, and there is virtually no difference between Eqs.~(\ref{eq:fulleqinspace1}). Also, note that the antineutrino flavor proportions discussed in Sec.~\ref{sec:2c} remain virtually unchanged for normal hierarchy while the neutrino flavor proportions change drastically around the point $\lambda_{r} \sim \omega_{r}$.
The characteristic oscillation length of the survival probability found on these plots is
\begin{equation}
\tau \approx (0.05-1) \,\,{\rm km}
\label{length}
\end{equation}
which agree with the ones given by Eq.~(\ref{deffinition tau}) calculated at the NS surface up to a factor of order one. Such a small value of $\tau$ suggests extremely quick $\nu_e\bar{\nu}_e \leftrightarrow \nu_x \bar{\nu}_x$ oscillations. 

Clearly, the full EoM are highly nonlinear so the solution may not reflect the real neutrino flavor evolution. Concerning the single-angle approximation, it is discussed in \citet{Hannestad:2006nj,Raffelt:2007yz,Fogli:2007bk} that in the more realistic multi-angle approach, kinematic decoherence happens. And in \citet{EstebanPretel:2007ec} the conditions for decoherence as a function of the neutrino flavor asymmetry have been discussed. It is concluded that if the symmetry of neutrinos and antineutrinos is broken beyond the limit of $O(25\%)$, i.e., if the difference between emitted neutrinos and antineutrinos is roughly larger than 25\% of the total number of neutrinos in the medium, decoherence becomes a sub-dominant effect. 

As a direct consequence of the peculiar symmetric situation we are dealing with, in which neutrinos and antineutrinos are produced in similar numbers, bipolar oscillations happen and, as we have already discussed, they present very small oscillation length as shown in Eq.~(\ref{length}). Note also that the bipolar oscillation length depends on the neutrino energy. Therefore, the resulting process is equivalent to an averaging over the neutrino energy spectrum and an equipartition among different neutrino flavors is expected~\cite{Raffelt:2007yz}. Although, for simplicity, we are dealing with the two neutrino hypothesis, this behavior is easily extended to the more realistic three neutrino situation. We assume, therefore, that at few kilometers from the emission region neutrino flavor equipartition is a reality:
\begin{equation}
\nu_e:\nu_\mu:\nu_\tau=1:1:1.
\label{eq:proportion}
\end{equation}

Note that the multi-angle approach keeps the order of the characteristic length $\tau$ of Eq.~(\ref{deffinition tau}) unchanged and kinematics decoherence happens within a few oscillation cycles \citep{Sawyer:2005jk,Hannestad:2006nj,Raffelt:2007yz}. Therefore, we expect that neutrinos created in regions close to the emission zone will be equally distributed among different flavors in less than few kilometers after their creation. Once the neutrinos reach this maximally mixed state, no further changes are expected up until the matter potential enters the MSW resonance region. We emphasize that kinematics decoherence does not mean quantum decoherence. Figs.~\ref{fig:singleangle} and \ref{fig:singleanglesolutions} clearly show the typical oscillation pattern which happens only if quantum coherence is still acting on the neutrino system. Differently from quantum decoherence, which would reveals itself by a monotonous dumping in the oscillation pattern, kinematics decoherence is just the result of averaging over the neutrino energy spectrum resulting from quick flavor conversion which oscillation length depends on the neutrino energy.  Therefore, neutrinos are yet able to quantum oscillate if appropriate conditions are satisfied. 

We discuss now the consequences of the matter potential.

\subsubsection{Matter Effects}\label{sec:mattereffects}

After leaving the emisison region, beyond $r\approx R_{\rm NS}+\Delta r_{\nu}$, where $\Delta r_{\nu}$ is the width defined in Eq.~(\ref{neutrinoshell}), the effective neutrino density quickly falls in a asymptotic behavior $\mu_{r} \approx 1/r^{4}$. The decay of $\lambda_{r}$ is slower. Hence, very soon the neutrino flavor evolution is determined by the matter potential. Matter suppresses neutrino oscillations and we do not expect significant changes in the neutrino flavor content along a large region. Nevertheless, the matter potential can be so small that there will be a region along the neutrino trajectory in which it can be compared with the neutrino vacuum frequencies and the higher and lower resonant density conditions will be satisfied, i.e.:
\begin{equation}
\lambda(r_H)=\omega_H=\frac{\Delta m^2}{2\langle E_{\nu} \rangle} {\hskip.2cm\mbox{and}\hskip.2cm} \lambda(r_L)=\omega_L=\frac{\Delta m^2_{21}}{2\langle E_{\nu} \rangle},
\label{resonances}
\end{equation}
where $\Delta m^2$ and $\Delta m^2_{21}$ are, respectively, the squared-mass differences found in atmospheric and solar neutrino observations. Table \ref{tab:tab3} shows the experimental values of mixing angles and mass-squared differences taken from \citet{Olive:2016xmw}. The definition of $\Delta m^2$ used is: $\Delta m^2 = m^2_3 -(m^2_2 + m^2_1)/2$. Thus, $\Delta m^2 = \Delta m^2_{31} - \Delta m^2_{21}/2 > 0$, if $m_1 < m_2 < m_3$, and $\Delta m^2 = \Delta m^2_{32} + \Delta m^2_{21}/2 < 0$ for $m_3 < m_1 < m_2$. When the above resonance conditions are satisfied the MSW effects happen and the flavor content of the flux of electronic neutrinos and antineutrinos will be again modified. The final fluxes can be written as
\begin{subequations}
\begin{gather}
F_{\nu_e}(E)=P_{\nu_e\to \nu_e}(E)F^0_{\nu_e}(E) +\left[1-P_{\nu_e\to \nu_e}(E)\right]F^0_{\nu_x}(E)\\
F_{\bar{\nu}_e}(E)=P_{\bar{\nu}_e\to \bar{\nu}_e}(E)F^0_{\bar\nu_e}(E) +\left[1-P_{\bar\nu_e\to \bar\nu_e}(E)\right]F^0_{\bar\nu_x}(E)
\end{gather}\label{eq:fluxatearth}
\end{subequations}
where $F^0_{\nu_e}(E)$, $F^0_{\nu_x}(E)$, $F^0_{\bar\nu_e}(E)$ and $F^0_{\bar\nu_x}(E)$ are the fluxes of electronic and non-electronic neutrinos and antineutrinos after the bipolar oscillations of the emission zone and $P_{\nu_e\to \nu_e}(E)$ and $P_{\bar\nu_e\to \bar\nu_e}(E)$ are the survival probability of electronic neutrinos and antineutrinos during the resonant regions.

In order to evaluate $F_{\nu_e}(E)$ and $F_{\bar\nu_e}(E)$ after matter effects, we have to estimate the survival probability at the resonant regions. There are several articles devoted to this issue; for instance we can adopt the result in \cite{Fogli:2003dw}, namely, for normal hierarchy
\begin{subequations}
\begin{gather}
P_{\nu_e\to \nu_e}(E)= X \sin^2\theta_{12}\\
P_{\bar\nu_e\to \bar\nu_e}(E)= \cos^2\theta_{12}
\end{gather}
\end{subequations}
and, for inverted hierarchy
\begin{subequations}
\begin{gather}
P_{\nu_e\to \nu_e}(E)=\sin^2\theta_{12}\\
P_{\bar\nu_e\to \bar\nu_e}(E)= X \cos^2\theta_{12}
\end{gather}
\end{subequations}
The factor $X$, the conversion probability between neutrino physical eigenstates, is given by \cite{Petcov:1987xd,Fogli:2003dw,Kneller:2005hf}
\begin{equation}
X = \frac{\exp(2πr_{\rm res}k_{\rm res} \cos 2 \theta_{13}) - 1}{\exp(2π r_{\rm res}k_{\rm res}) - 1},
 \label{eqX}
 \end{equation}
where $r_{\rm res}=r_L$ or $r_{\rm res}=r_H$, defined according to Eq.~(\ref{resonances}) and  
\begin{equation}
\frac{1}{k_{\rm res}} = \left\vert \frac{d \ln \lambda _{r}}{dx}\right\vert_{r=r_{\rm res}}.
\label{eqk}
\end{equation}
The factor $X$ is related to how fast physical environment features relevant for neutrino oscillations change, such as neutrino and matter densities.

For slow and adiabatic changes $X\rightarrow 0$ while for fast and non-adiabatic, $X\rightarrow 1$. In our specific cases, the MSW resonances occur very far from the accretion zone where the matter density varies very slow and therefore $X \rightarrow 0$, as can be explicitly calculated from Eq.~(\ref{eqX}). Consequently, it is straightforward to estimate the final fluxes of electronic and non-electronic neutrinos and antineutrinos.

\subsection{Low accretion rates}\label{sec:lowac}

%
\begin{figure*}
\centering
\includegraphics[width=0.44\hsize,clip]{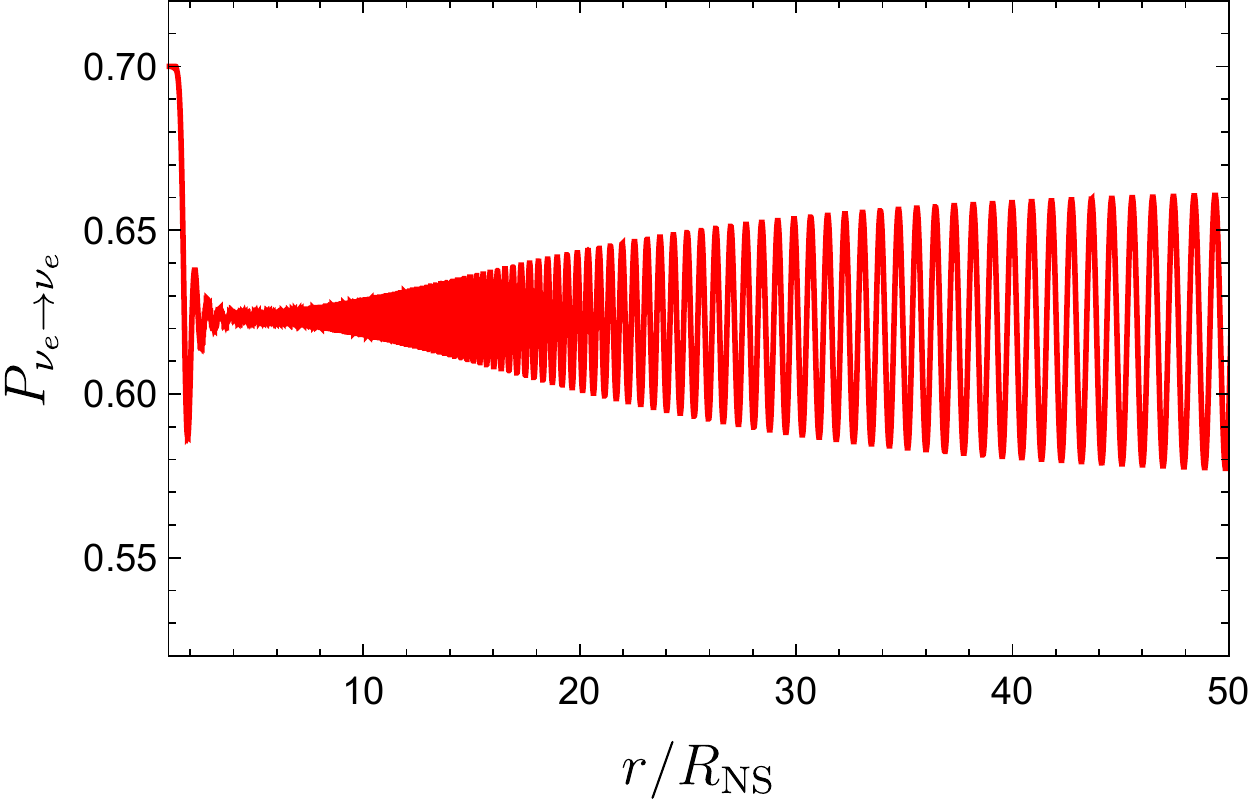}\includegraphics[width=0.44\hsize,clip]{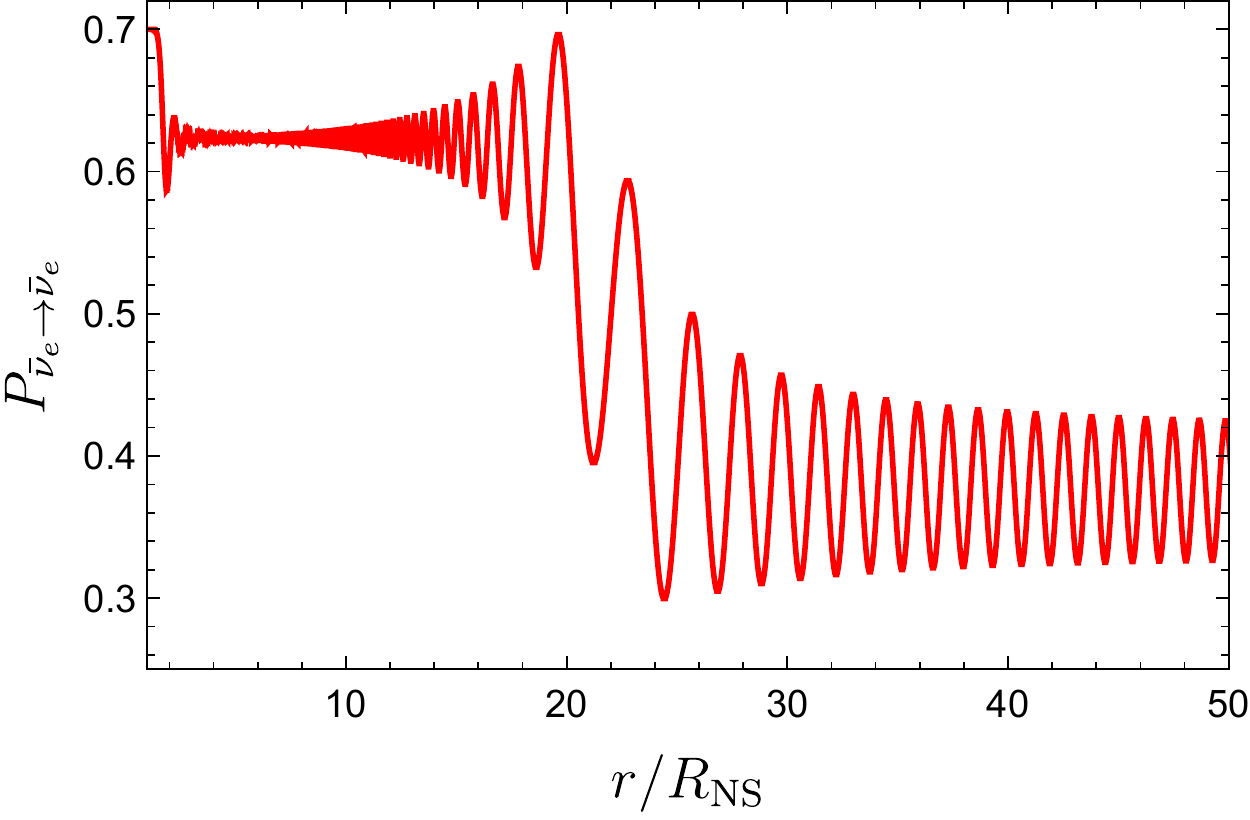}
\caption{Electron neutrino and antineutrino flavor evolution for inverted hierarchy and $\dot{M}=10^{-6}M_{\odot}$~s$^{-1}$. The survival probability is shown as a function of the radial distance from the NS surface.} 
\label{fig:examplelowac}
\end{figure*}

For accretion rates $\dot{M} < 5\times\!10^{-5}M_{\odot}$~s$^{-1}$, either the matter potential is close enough to the vacuum potential and the MSW condition is satisfied, or both the self-interaction and matter potentials are so low that the flavor oscillations are only due to the vacuum potential. In both cases, bipolar oscillations are not present. In Fig.~(\ref{fig:examplelowac}) we show the survival probability for $\dot{M}=10^{-6}M_{\odot}$~s$^{-1}$ as an example. We can see that neutrinos and antineutrinos follow different dynamics. In particular, for antineutrinos there are two decreases. The first one, around $r \approx (1$--$2) R_{\rm NS}$, is due to bipolar oscillations which are rapidly damped by the matter potential as discussed in Sec.~\ref{sec:mattereffects}. The second one happens around $r \approx (10$--$20) R_{\rm NS}$. It can be seen from the bottom left panel of Fig.~\ref{fig:potentials} (that one for $\dot{M} = 10^{-6}M_{\odot}$~s$^{-1}$), that around $r \approx (1$--$2) \times 10^{7}$~cm (or, equivalently, $r \approx(10$--$20)\, R_{\rm NS}$) the higher MSW resonance occurs ($\lambda_{r} \sim \omega_{r_H}$). For inverted hierarchy, such resonance will affect antineutrinos depleting its number, as can be seen from Eq.~(\ref{eq:fluxatearth}). Without bipolar oscillations, it is not possible to guarantee that decoherence will be complete and Eq.~(\ref{eq:proportion}) is no longer valid. The only way to know the exact flavor proportions is to solve the full Eqs.~(\ref{eq:Hnu}).

\begin{table*}
\centering
\begin{tabular}{c c c c c c c c c c c c}
\hline
  & $n^{0}_{\nu_{e}}/n$\T\B & $n^{0}_{\bar{\nu}_{e}}/n$\T\B & $n^{0}_{\nu_{x}}/n$\T\B & $n^{0}_{\bar{\nu}_{x}}/n$\T\B & $n_{\nu_{e}}/n$\T\B & $n_{\bar{\nu}_{e}}/n$\T\B & $n_{\nu_{x}}/n$\T\B & $n_{\bar{\nu}_{x}}/n$\T\B \\ 
 \hline\hline
    Normal Hierarchy\T\B & $\frac{1}{6}$\T & $\frac{1}{6}$\T & $\frac{1}{3}$\T & $\frac{1}{3}$\T & $\frac{1}{3}$\T & $\frac{1}{6} + \frac{1}{6}\sin^{2}\theta_{12}$\T & $\frac{1}{6}$\T & $\frac{1}{3}-\frac{1}{6}\sin^{2}\theta_{12}$\T \\ \hline
    Inverted Hierarchy\T\B & $\frac{1}{6}$\T & $\frac{1}{6}$\T & $\frac{1}{3}$\T & $\frac{1}{3}$\T & $\frac{1}{6} + \frac{1}{6}\cos^{2}\theta_{12}$\T & $\frac{1}{3}$\T & $\frac{1}{3}-\frac{1}{6}\cos^{2}\theta_{12}$\T & $\frac{1}{6}$\T \\ \hline    
\end{tabular}
\caption{Fraction of neutrinos and antineutrinos for each flavor after decoherence and matter effects. $n=2\sum_{i}n_{\nu_{i}}$.}
\label{tab:tabfluxes}
\end{table*}
\section{Neutrino Emission spectra}\label{sec:spectra}

Using the the calculations of last section we can draw a comparison between the creation spectra of neutrinos and antineutrinos at the NS surface ($F^c_\nu,n^c_\nu$), initial spectra after kinematic decoherence ($F^0_\nu,n^0_\nu$) and emission spectra after the MSW resonances ($F_\nu,n_\nu$). Table~\ref{tab:tabfluxes} contains a summary of the flavor content inside the Bondi-Hoyle radius. With these fractions and Eqs.~(\ref{eq:neutrinofullinitialspectrum}) it is possible to reproduce the spectrum for each flavor and for accretion rates $M \geq 5 \!\times\! 10^{-5}M_{\odot}$~s$^{-1}$.

\begin{figure*}
\centering
\includegraphics[width=0.49\hsize,clip]{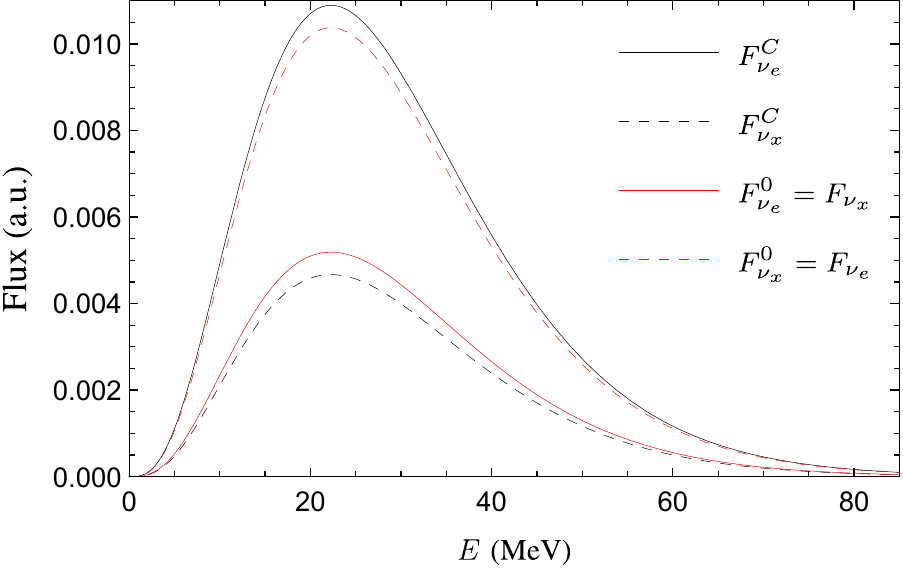}\includegraphics[width=0.49\hsize,clip]{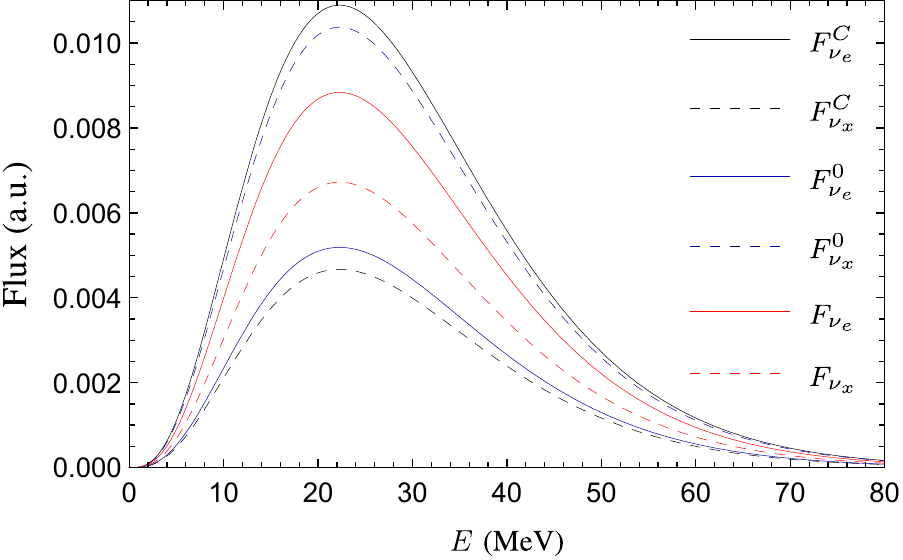}\\
\includegraphics[width=0.49\hsize,clip]{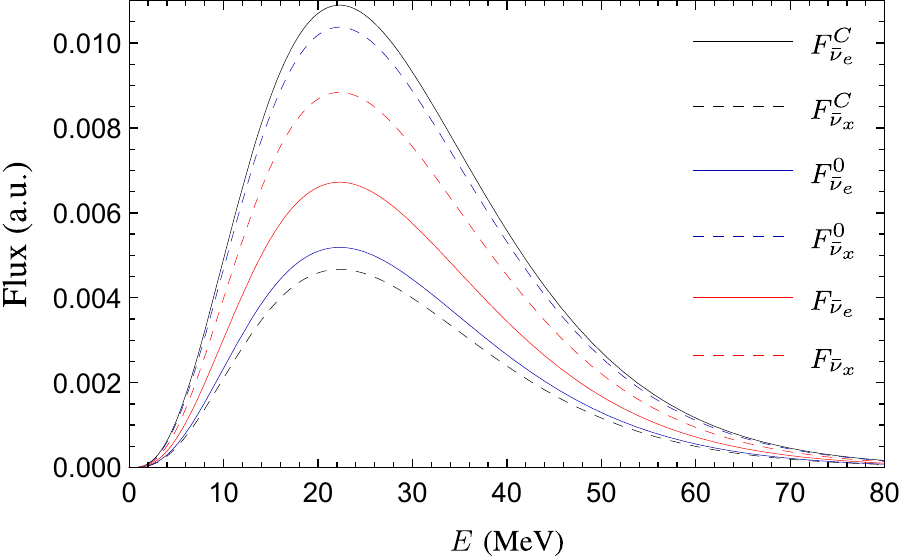}\includegraphics[width=0.49\hsize,clip]{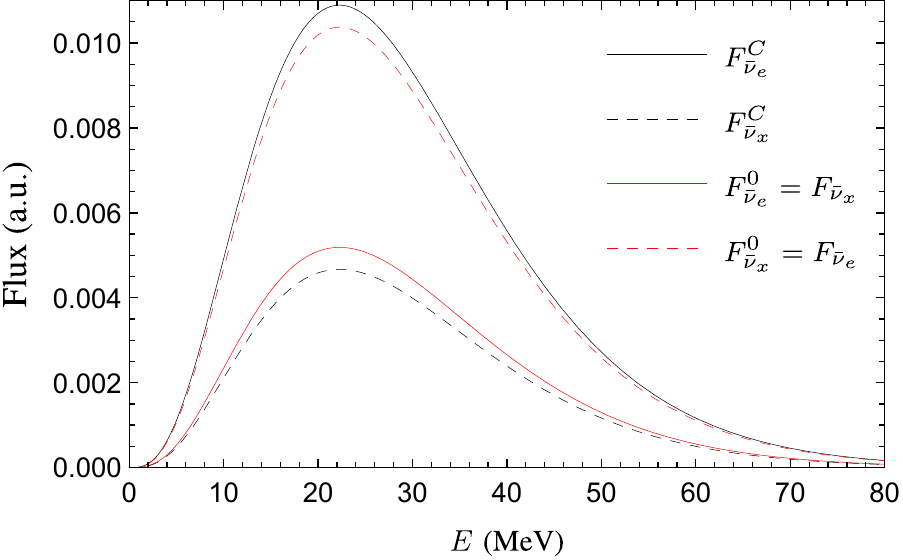}\\
\includegraphics[width=0.49\hsize,clip]{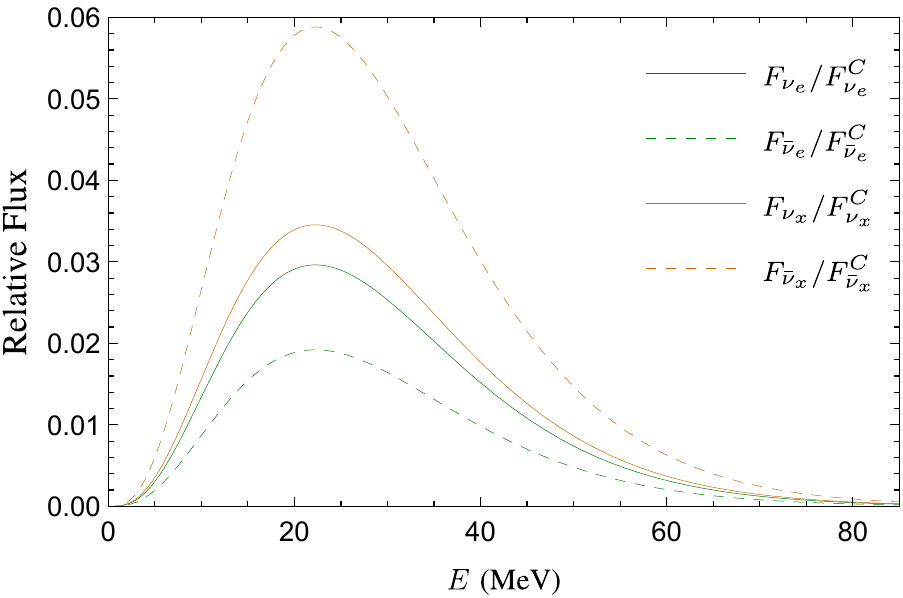}\includegraphics[width=0.49\hsize,clip]{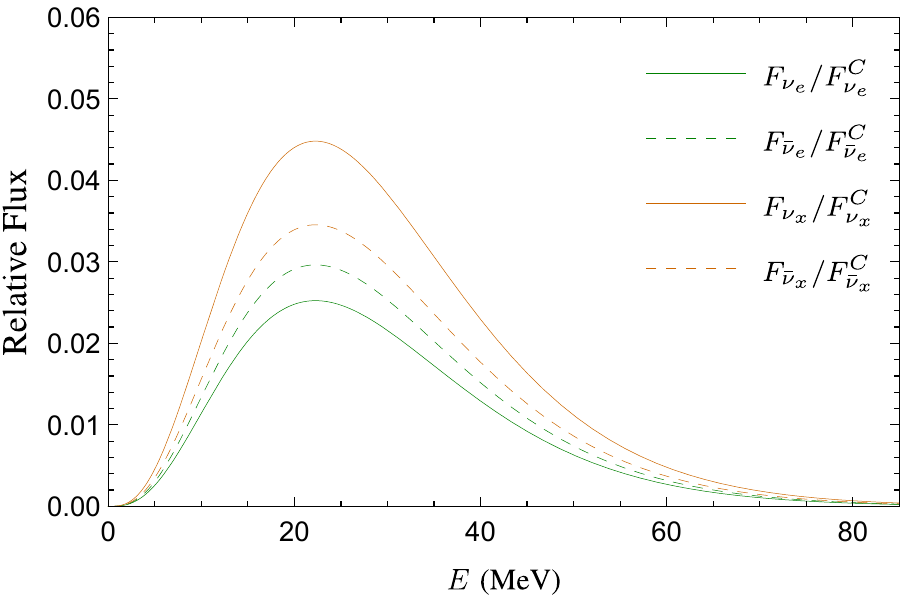}\caption{Several neutrino and antineutrino number fluxes for different neutrino flavors are presented for $\dot{M}=10^{-2}M_{\odot}/{\rm s}$. Each column corresponds to a neutrino mass hierarchy: normal hierarchy on the left and inverted hierarchy on the right. The first two rows show the number fluxes after each process studied. $F^{C}_{\nu}$, $F^{0}_{\nu}$ and $F_{\nu}$ are the creation flux at the bottom accretion zone due to $e^{+}e^{-}$ pair annihilation, the flux after the region with dominant neutrino-neutrino potential and the final emission flux after the region with dominant neutrino-matter potential, respectively. The last row shows the relative fluxes $F_{\nu}/F^{C}_{\nu}$ between the creation and emission fluxes.} 
\label{fig:EmissionTotal}
\end{figure*}

The specific cases for $\dot{M}=10^{-2}M_{\odot}$~s$^{-1}$ are shown in Fig.~\ref{fig:EmissionTotal}. In such figures, the left column corresponds to normal hierarchy and the right corresponds to inverted hierarchy. The first two rows show the number fluxes after each process studied. The last row shows the relative fluxes $F_{\nu}/F^{C}_{\nu}$ between the creation and emission fluxes.  For the sake of clarity, we have normalized the curves to the total neutrino number at the NS surface
\begin{equation}
n=2\!\sum_{i\in\{e,x\}}\!n_{\nu_{i}}.
\label{eq:totalnumber}
\end{equation}
so that each one is a normalized Fermi-Dirac distribution multiplied by the appropriate flavor content fraction. To reproduce any other case, it is enough to use Eqs.~(\ref{eq:neutrinofullinitialspectrum}) with the appropriate temperature.

At this point two comments have to be made about our results:

\begin{itemize}
\item As we mentioned before, the fractions in Table~\ref{tab:tabfluxes} were obtained by assuming a monochromatic spectrum and using the single-angle approximation. This would imply that the spectrum dependent phenomenon called the \emph{spectral stepwise swap} of flavors is not present in our analysis even though it has been shown that it can also appear in multi-angle simulations \citep{Fogli:2007bk}. Nevertheless, we know from our calculations in Sec.~\ref{sec:2c} that neutrinos and antineutrinos of all flavors are created with the exact same spectrum up to a multiplicative constant. Hence, following \cite{Raffelt:2007xt,Raffelt:2007cb}, by solving the equation
\begin{equation}
\int^{\infty}_{E_{c}}\left( n_{\nu_{e}} - n_{\nu_{x}} \right)dE=\int^{\infty}_{0}\left( n_{\bar{\nu}_{e}} - n_{\bar{\nu}_{x}} \right)dE,
\label{eq:energysplit}
\end{equation}
we find that the critical (split) energy is $E_{c} = 0$. This means that the resulting spectrum should still be unimodal and the spectral swap in our system could be approximated by a multiplicative constant that is taken into account in the decoherence analysis of Sec.~\ref{sec:solutions}.

\item The fluxes of electronic neutrinos and antineutrinos shown in these figures and in Eqs.~(\ref{eq:fluxatearth}) represent fluxes at different positions up to a geometrical $1/r^2$ factor, $r$ being the distance from the NS radius. Also, since we are considering the fluxes before and after each oscillatory process, the values of $r$ are restricted to $r=R_{\rm NS}$ for $F^{C}_{\nu}$, $\tau_{\dot{M}} < r < r_{H}$ for $F^{0}_{\nu}$, and $r > r_{L}$ for $F_{\nu}$. To calculate the number flux at a detector, for example, much higher values of $r$ have to be considered and it is necessary to study vacuum oscillations in more detail. Such calculations will be presented elsewhere.

\end{itemize}

From Fig.~\ref{fig:EmissionTotal} one can observe that the dominance of electronic neutrinos and antineutrinos found at their creation at the bottom of the accretion zone is promptly erased by kinematic decoherence in such a way that the content of the neutrinos and antineutrinos entering the MSW resonant region is dominated by non-electronic flavors. After the adiabatic transitions provoked by MSW transitions, electronic neutrinos and antineutrinos dominate again the emission spectrum except for non-electronic antineutrinos in the normal hierarchy.  Although no energy spectrum distortion is expected, the flavor content of neutrinos and antineutrinos produced near the NS surface escape to the outer space in completely different spectra when compared with the ones in which they were created, as shown in the last row of Fig.~\ref{fig:EmissionTotal}.

\section{Concluding remarks}\label{sec:4}

We can now proceed to draw the conclusions and some astrophysical consequences of this work:
\begin{enumerate}

\item
The main neutrino production channel in XRFs and BdHNe in the hypercritical accretion process is pair annihilation: $e^{-}\!e^{+}\!\rightarrow\nu\bar\nu$. This mechanism produces an initial equal number of neutrino and antineutrino and an initial 7/3 relative fraction between electronic and other flavors. These features lead to a different neutrino phenomenology with respect to the typical core-collapse SN neutrinos produced via the URCA process.

\item
The neutrino density is higher than both the electron density and the vacuum oscillation frequencies for the inner layers of the accretion zone and the self-interaction potential dictates the flavor evolution along this region, as it is illustrated by Fig.~\ref{fig:potentials}. This particular system leads to very fast pair conversions $\nu_e\bar{\nu}_e\!\!\leftrightarrow\!\! \nu_{\mu,\tau}\bar{\nu}_{\mu, \tau}$ induced by bipolar oscillations with oscillation length as small as $O(0.05$--$1)$~km. However, due to the characteristics of the main neutrino production process, neutrinos and antineutrinos have very similar fluxes inside the neutrino emission zone and kinematic decoherence dominates the evolution of the polarization vectors.

\item
The kinematic decoherence induces a fast flux equipartition among the different flavors that then enters the matter dominated regions in which MSW resonances take place.

\item
Therefore, the neutrino flavor content emerging from the Bondi-Hoyle surface to the outer space is different from the original one at the bottom of the accretion zone. As shown in Table~\ref{tab:tabfluxes}, The initial 70\% and 30\% distribution of electronic and non-electronic neutrinos becomes 55\% and 45\% or 62\% and 38\% for normal or inverted hierarchy, respectively. Since the $\nu\leftrightarrow\bar{\nu}$ oscillations are negligible \citep{Pontecorvo:1957cp,Pontecorvo:1967fh,Xing:2013ty} the total neutrino to antineutrino ratio is kept constant.

\end{enumerate}

We have shown that such a rich neutrino phenomenology is uniquely present in the hypercritical accretion process in XRFs and BdHNe. This deserves the appropriate attention since it paves the way for a new arena of neutrino astrophysics besides SN neutrinos. There are a number of issues which have still to be investigated:
\begin{enumerate}

\item We have made some assumptions which, albeit being a first approximation to a more detailed picture, have allowed us to set the main framework to analyze the neutrino oscillations phenomenology in these systems. We have shown in \citet{2015ApJ...812..100B} that the SN ejecta carry enough angular momentum to form a disk-like structure around the NS before being accreted. However, the knowledge of the specific properties of such possible disk-like structure surrounding the neutron star is still pending of more accurate numerical simulations at such distance scales. For instance, it is not clear yet if such a structure could be modeled via thin-disk or thick-disk models. We have adopted a simplified model assuming isotropic accretion and the structure of the NS accretion region used in \citet{2016ApJ...833..107B} which accounts for the general physical properties of the system. In order to solve the hydrodynamics equations, the neutrino-emission region features, and the neutrino flavor-oscillation equations, we have assumed: spherically symmetric accretion onto a non–rotating NS, a quasi-steady-state evolution parametrized by the mass accretion rate, a polytropic equation of state, and subsonic velocities inside the shock radius. The matter is described by a perfect gas made of ions, electrons, positrons and radiation with electron and positron obeying a Fermi-Dirac distribution. The electron fraction was fixed and equal to 0.5. We considered pair annihilation, photo-neutrino process, plasmon decay and bremsstrahlung to calculate neutrino emissivities. Under the above conditions we have found that the pair annihilation dominates the neutrino emission for the accretion rates involved in XRFs and BdHNe \citep[see][for further details]{2016ApJ...833..107B}. The photons are trapped within the infalling material and the neutrinos are transparent, taking away most of the energy from the accretion. We are currently working on the relaxation of some of the above assumptions, e.g. the assumption of spherical symmetry to introduce a disk-like accretion picture, and the results will be presented elsewhere. In this line it is worth mentioning that some works have been done in this direction \citep[see, e.g.,][]{Zhang:2007ds,2009ApJ...703..461Z}, although in a Newtonian framework, for complete dissociated matter, and within the thin-disk approximation. In these models, disk heights $H$ are found to obey the relation $H/r \sim 0.1$ near the neutron star surface which suggests that the results might be similar to the ones of a spherical accretion as the ones we have adopted. We are currently working on a generalization including general relativistic effects in axial symmetry to account for the fast rotation that the NS acquires during the accretion process. This was already implemented for the computation of the accretion rates at the Bondi-Hoyle radius position in \citet{2016ApJ...833..107B}, but it still needs to be implemented in the computation of the matter and neutrino density-temperature structure near the NS surface. In addition, the description of the equation of state of the infalling matter can be further improved by taking into account beta and nuclear statistical equilibrium.

In forthcoming works we will relax the assumptions made not only on the binary system parameters but also make more detailed calculations on the neutrino oscillations including general relativistic and multi-angle effects. This paper, besides presenting a comprehensive non-relativistic account of flavor transformations in spherical accretion, serves as a primer that has allowed us to identify key theoretical and numerical features involved in the study of neutrino oscillations in the IGC scenario of GRBs. From this understanding, we can infer that neutrino oscillations might be markedly different in a disk-like accretion process. First, depending on the value of the neutron-star mass, the inner disk radius may be located at an $r_{\rm inner}>R_{\rm NS}$ beyond the NS surface \citep[see e.g.][]{2016ApJ...832..136R,2017PhRvD..96b4046C}, hence the  neutrino emission must be located at a distance $r\geq r_{\rm inner}$. On the other hand, depending on the accretion rate, the density near the inner radius can be higher than in the present case and move the condition for neutrino cooling farther from the inner disk radius, at $r>r_{\rm inner}$. Both of these conditions would change the geometric set up of the neutrino emission. Furthermore, possible larger values of $T$ and $\rho$ may change the mechanisms involved in neutrino production. For example, electron-positron pair capture, namely $p+e^{-} \rightarrow n+\nu_{e}$, $n+e^{+} \rightarrow p+\bar{\nu}_{e}$ and $n \rightarrow p+e^{-}+\bar{\nu}_{e}$, may become as efficient as the electron-positron pair annihilation. This, besides changing the intensity of the neutrino emission, would change the initial neutrino-flavor configuration.

\item
Having obtained the flux as well as the total number of neutrinos and antineutrinos of each flavor that leave the binary system during the hypercritical accretion process in XRFs and BdHNe, it raises naturally the question of the possibility for such neutrinos to be detected in current neutrino observatories. For instance, detectors such as Hyper-Kamiokande are more sensitive to the inverse beta decay events produced in the detector, i.e. $\bar{\nu}_e + p \to e^+ + n$ \citep[see][for more details]{2011arXiv1109.3262A}, consequently, the $\bar{\nu}_e$ are the most plausible neutrinos to be detected. \citet{2016PhRvD..93l3004L} have pointed out that for a total energy in $\bar{\nu}_e$ of $10^{52}$~erg and $\langle E_{\bar{\nu}_e}\rangle\sim 20$~MeV, the Hyper-Kamiokande neutrino-horizon is of the order of 1~Mpc. In the more energetic case of BdHNe we have typically $\langle E_{\nu,\bar\nu}\rangle\sim 20$~MeV (see table \ref{tab:tab1}) and a total energy carried out $\bar{\nu}_e$ of the order of the gravitational energy gain by accretion, i.e. $E_g\sim 10^{52}$--$10^{53}$~erg. Therefore we expect the BdHN neutrino-horizon distance to be also of the order of $1$~Mpc. These order-of-magnitude estimates need to be confirmed by detailed calculations, including the vacuum oscillations experienced by the neutrinos during their travel to the detector, which we are going to present elsewhere.
\item If we adopt the local BdHNe rate $\sim 1$~Gpc$^{-3}$~yr$^{-1}$ \citep{2016ApJ...832..136R} and the data reported above at face value, it seems that the direct detection of this neutrino signal is very unlikely. However, the physics of neutrino oscillations may have consequences on the powering mechanisms of GRBs such as the electron-positron pair production by neutrino-pair annihilation. The energy deposition rate of this process depends on the local energy-momentum distribution of (anti)neutrinos which, as we have discussed, is affected by the flavor oscillation dynamics. This phenomenon may lead to measurable effects on the GRB  emission.
\item
An IGC binary leading either to an XRF or to a BdHN is a unique neutrino-physics laboratory in which there are at least three neutrino emission channels at the early stages of the GRB-emission process: i) the neutrinos emitted in the explosion of the CO$_{\rm core}$ as SN; ii) the neutrinos studied in this work created in the hypercritical accretion process triggered by the above SN onto the NS companion, and iii) the neutrinos from fallback accretion onto the $\nu$NS created at the center of the SN explosion. It remains to establish the precise neutrino time sequence as well as the precise relative neutrino emissivities from all these events. This is relevant to establish both the time delays in the neutrino signals as well as their fluxes which will become a unique signature of GRB neutrinos following the IGC paradigm.

\item
As discussed in \citet{2016ApJ...832..136R}, there are two cases in which there is the possibility to have hypercritical accretion onto a BH. First, in BdHNe there could be still some SN material around the newly-born BH which can create a new hypercritical accretion process \citep{2016ApJ...833..107B}. Second, a $\sim10~M_\odot$ BH could be already formed before the SN explosion, namely the GRB could be produced in a CO$_{\rm core}$-BH binary progenitor. The conditions of temperature and density in the vicinity of these BHs might be very different to the ones analyzed here and, therefore, the neutrino emission and its associated phenomenology. We have recalled in the introduction that such an accretion process onto the BH can explain the observed 0.1--100~GeV emission in BdHNe (\citealp{2015ApJ...798...10R,2015ApJ...808..190R,2016ApJ...831..178R,2016ApJ...832..136R,2017ApJ...844...83A}; see also Aimuratov et al. in preparation). The interaction of such an ultra-relativistic expanding emitter with the interstellar medium could be a possible source of high-energy (e.g. TeV-PeV) neutrinos, following a mechanisms similar to the one introduced in the traditional collapsar-fireball model of long GRBs \citep[see e.g.][and references therein]{2017APh....86...11A,2015PhR...561....1K}.

\item
Although the symmetry between the neutrino and antineutrino number densities has allowed us to generalize the results obtained within the single-angle and monochromatic spectrum approximations, to successfully answer the question of detectability, full-scale numerical solutions will be considered in the future to obtain a precise picture of the neutrino-emission spectrum. In particular, it would be possible to obtain an $r$-dependent neutrino spectrum without the restrictions discussed in Sec.~\ref{sec:spectra}.

\item
For low accretion rates ($\dot{M} \lesssim 5\times10^{-5}\,M_{\odot}$~s$^{-1}$) the matter and self-interaction potentials in Eqs.~(\ref{eq:fulleqinspace1}) decrease and the general picture described in Fig.~\ref{fig:potentials} changes. The resonance region could be located around closer to the NS surface, anticipating the MSW condition $\lambda_{r}\sim\omega_{r}$ and interfering with the kinematic decoherence. This changes the neutrino flavor evolution and, of course, the emission spectrum. Hence, the signature neutrino-emission spectrum associated with the least luminous XRFs might be different from the ones reported here.
\end{enumerate}

\acknowledgments

We thank the Referee for the comments and suggestions which helped to make more clear the presentation of our results. M.M.G. thanks FAPESP (contract number 2016/00799--7) for the financial support and ICRANet in Pescara and Rome for the hospitality during the realization of this article. R.R. acknowledges the collaboration ICRANet-INFN. J.D.U. thanks COLCIENCIAS for the financial support. J.A.R acknowledges the partial support of the project N. 3101/GF4 IPC-11, and the target program F.0679 of the Ministry of Education and Science of the Republic of Kazakhstan. 

\bibliographystyle{apj}
\bibliography{main}

\end{document}